\documentclass[%
 aip,
 jcp,%
 amsmath,amssymb,
 reprint,%
final,
]{revtex4-1}
\usepackage{graphicx}
\usepackage{dcolumn}
\usepackage{bm}
\usepackage{color}
\usepackage{xcolor}
\usepackage{subcaption}
\usepackage{tikz}

\usepackage{calc}
\newlength{\charwidth}
\newlength{\inda}
\newlength{\indaa}
\newlength{\indb}
\newlength{\indbb}
\newlength{\indc}
\newlength{\indcc}
\newlength{\indd}
\newlength{\inddd}
\begin{document}


\title{Computationally Efficient Characterization of Potential Energy Surfaces Based on Fingerprint Distances} 


\author{Bastian Schaefer}
\affiliation{Department of Physics, University of Basel,
Klingelbergstrasse 82, CH-4056 Basel, Switzerland}

\author{Stefan Goedecker} \email[]{stefan.goedecker@unibas.ch}
\affiliation{Department of Physics, University of Basel,
Klingelbergstrasse 82, CH-4056 Basel, Switzerland}


\date{\today}

\begin{abstract}
An analysis of the network defined by the potential energy minima of
multi-atomic systems and their connectivity via reaction pathways that go
through transition states allows to understand important characteristics like
thermodynamic, dynamic and structural properties.  Unfortunately computing the
transition states and reaction pathways in addition to the significant
energetically low-lying local minima is a computationally demanding task.  We
here introduce a computationally efficient method that is based on a
combination of the minima hopping global optimization method and the insight
that uphill barriers tend to increase with increasing structural distances of
the educt and product states. This method allows to replace the exact
connectivity information and transition state energies with alternative and
approximate concepts.  Without adding any significant additional cost to the
minima hopping global optimization approach, this method allows to generate an
approximate network of the minima, their connectivity and a rough measure for
the energy needed for their interconversion. This can be used to obtain a first
qualitative idea on important physical and chemical properties by means of a
disconnectivity graph analysis.  Besides the physical insight obtained by such
an analysis, the gained knowledge can be used to make a decision if it is
worthwhile or not to invest computational resources for an exact computation of
the transition states and the reaction pathways.  Furthermore it is
demonstrated that the here presented method can be used for finding physically
reasonable interconversion pathways that are promising input pathways for
methods like transition path sampling or discrete path sampling.
\end{abstract}


\maketitle 

\section{Introduction}\label{Introduction}
Thermodynamic and kinetic
properties of multi-atomic systems are encoded
in the topology of their potential energy surfaces (PES). For example,
the folding of a protein into its native state seems to be impossible
based on the sheer abundance of conformational possibilities
(Levinthal's paradox).\cite{Levinthal1969} However, a steep
funnel-like shape of the PES results in driving forces that rapidly
lead the system towards its stable configuration, independent of its
initial denatured structure.\cite{Dill1997} In contrast, multi-funnel
PES can explain why a certain system might be observed in a metastable
state. Glass formation can be identified with trapping in some
disordered state.\cite{Wales2003} Accurately assessing the shape of a
PES usually requires not only the computation of local minima, but also
the network of possible transitions and the corresponding energy
barriers.

There exist various methods such as transition path sampling
(TPS),\cite{Dellago1998-1,Dellago1998-2,Bolhuis2002,Dellago2003,Gruenwald2008,Gruenwald2009,Lechner2011}
discrete path sampling (DPS),\cite{Wales2002,Wales2004-2} stochastic surface
walking based reaction sampling (SSW-RS),\cite{Zhang2015} the activation
relaxation technique nouveau
(ARTn),\cite{Mousseau1998,Wei2002,Machado2011,Mousseau2012} temperature
accelerated dynamics (TAD),\cite{Sorensen2000,Perez2009} or the minima hopping
guided path sampling (MHGPS)
approach,\cite{Goedecker2004,Schaefer2014,Schaefer2015} that allow the rigorous
sampling of reactive processes. Some of these methods can be even used at
sophisticated levels of theory, like, for example, at the density functional
(DFT) level. Nevertheless, these methods are computationally very demanding,
typically even more costly than the already challenging global
optimization\cite{Goedecker2004,Amsler2010,Mousseau1998,Wei2002,Machado2011,Mousseau2012,Oganov2006,Glass2006,Wales1997,Doye1998,Doye1998b,Pickard2006,Pickard2007,Pickard2007b,Pickard2008}
problem.  Therefore, computationally lightweight methods that allow to obtain
at least a qualitative impression of a PES are of high interest. To this end we
recently introduced distance-energy (DE) plots that allow to distinguish glassy
from non-glassy systems.\cite{De2014} In a DE plot the (atomization) energies
per atom of metastable configurations are measured relatively to the global
minimum and they are plotted versus their configurational distance to the
global minimum.  Structural fingerprints, which are based on the overlap matrix
of Gaussian type orbitals, can be used for measuring the configurational
distances.\cite{De2014,Sadeghi2013} Even on demanding levels of theory like
DFT, it is computationally feasible to produce DE plots, because only the
geometries and energies of a few hundred energetically low-lying local minima,
including the global minimum, are needed. 


In contrast to the disconnectivity graphs of Becker and
Karplus,\cite{Becker1997,Wales2003} DE plots contain different and
complementary information. DE plots focus on the relation of metastable
configurations to the global minimum and display the density of the
structures both with respect to energies and with respect to
configurational distances. This allows the deduction of a measure for
the driving force towards the global minimum. However, DE plots give no
relation between two arbitrary minima and, therefore, cannot display
topological information beyond the driving force towards the global
minimum. This is a consequence of the very modest requirements of DE
plots: only the energies and geometries of the global minimum and a few
hundred energetically low-lying local minima are needed. There is no need for transition state
energies or the information, which minima are connected with each other
by only one intermediate transition state.  However, in this
contribution it is demonstrated that, based only on the data obtained
during conventional MH runs, an approximation to this connectivity
information is available.  Furthermore, it is discussed that an
empirical guess for the transition state energies can be obtained,
which is based solely on fingerprint distances of local minima. The
combination of the approximate connectivity information and the guess
for the transition state energies allows to generate a new type of
disconnectivity graph that shows a remarkable resemblance to
disconnectivity graphs which are based on exact transition state
energies and exact connectivity information.  The post-processing of
the MH data for the generation of DE plots, for the extraction of the
approximate connectivity information and for the computation of the
transition state energy guess can conveniently be performed on a single
core of a standard personal computer within a negligible amount of
wall-clock time.  Therefore, DE plots and the method presented in this
contribution give a useful and computationally very affordable overview of the
characteristics of a PES. They can serve as a valuable aid for making a
decision whether investing the computer        time that is required
for building a rigorous network of transitions and their corresponding
barrier energies is worthwhile and expedient with respect to a certain
research goal, or not.                     Furthermore, they provide a
qualitative idea on the kinetics       and thermodynamics of a system.
Moreover, the method presented below is demonstrated to be a promising
tool for isolating physically reasonable intermediate metastable
structures of complicated reactions, which, for example, might be used
for generating initial pathways that are needed in methods like TPS or
its discrete variant, DPS.

\section{Correlating Transition State Energies with Structural
Differences}\label{sec:corrTSener}
\begin{figure}[t]
\centering
\includegraphics[scale=0.8]{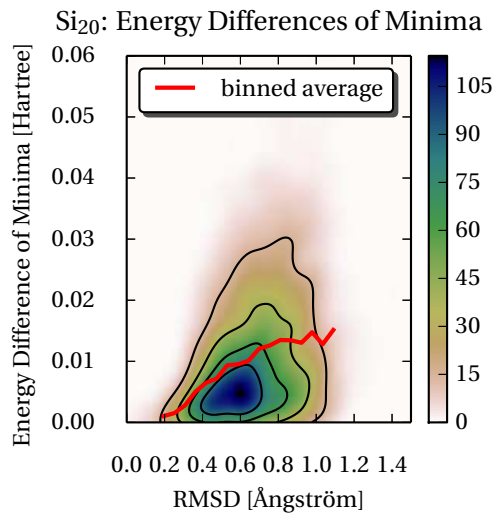}
  \caption{\label{fig:si20minenerdiff} Density plot of the energy
differences of pairs of minima versus their RMSD distance for the $Si_{\mbox{20}}$ system. The shown
data sets consists of roughly 2900 minima pairs. Each pair of minima is
connected by only one intermediate transition state. The structures,
energies and the connectivity of the stationary points were determined
at the DFT level of theory (PBE exchange correlation functional) by
using the MHGPS method coupled to the BigDFT
code.\cite{Genovese2008,Mohr2014,Willand2013,Schaefer2014,Schaefer2015} The shown
density was obtained from the corresponding scattered data by means of
a Gaussian kernel density estimate as implemented in Python's scipy
library. The red bold line shows the same data, but averaged within 25
bins along the RMSD axis.  Only bins that contain at least 5\% of the
number of data points of the bin with the most data points are shown.}
\end{figure}
\begin{figure*}
\centering
\includegraphics{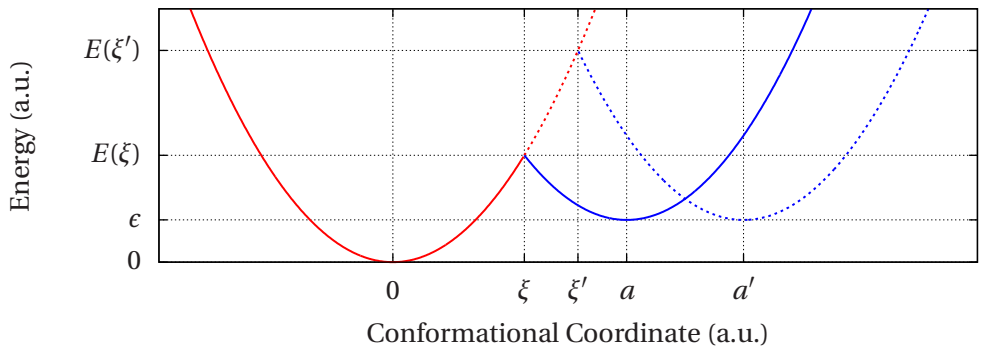}
\caption{\label{fig:rmsdbemodel} Parabola model for the transition state
energy.
For increasing structural differences of both minima the transition state
energy is rising. Here this is illustrated by shifting the minimum of the solid
blue parabola from $a$ to $a'$. The sifted parabola is visualized by a blue
dashed line.}
\end{figure*}

Often the energies of two structurally similar minima of a PES are very
close to each other, whereas the energy differences between
structurally very different minima can be large. Nevertheless, it is
clear that structurally very different minima can have very similar
energies, as well. In other words, it is expected that for small
structural differences the probability to find large energy differences
is small, whereas for large structural differences, both, small and
large energy differences between two adjacent minima are likely.
Indeed, this expectation is supported by the data shown in
Fig.~\ref{fig:si20minenerdiff}. For the neutral silicon cluster
consisting of 20 atoms, this figure shows the density of the
distribution of energy differences of minima pairs plotted versus the
corresponding permutationally optimized RMSD
distance.\cite{Sadeghi2013} All minima pairs used for this plot are
separated by only one intermediate transition state. It is seen from
this plot, that for small RMSD values the density of the data points
vanishes for large energy differences, whereas for larger RMSD values,
there is a significant density, both for small and large energy
differences.  Because the variance in the energy differences is larger
for increasing RMSD values, also the average values of the energy
differences rises, as is shown by the binned average of the energy
differences (red line).

Except for degenerate rearrangements, the barrier energy of every transition
state can be measured with respect to the lower or the higher energy minimum to
which the transition state is connected to.  In contrast to the distribution of
the energy differences of neighboring minima in an energy difference versus
RMSD plot, it can be expected that there is a stronger correlation in a plot of
the uphill (larger) barriers versus the RMSD. Intuitively, this partially
should result from a combination of the fact that the absolute values of the
energy differences of two neighboring minima are a lower bound for the uphill
barriers and the assumption that the average downhill barrier energy should
rise if the distance between the minima increases.  Therefore, the probability
to find small uphill barriers between structurally very different minima should
be expected to be small.

\begin{figure}
\centering
    \includegraphics[scale=0.8]{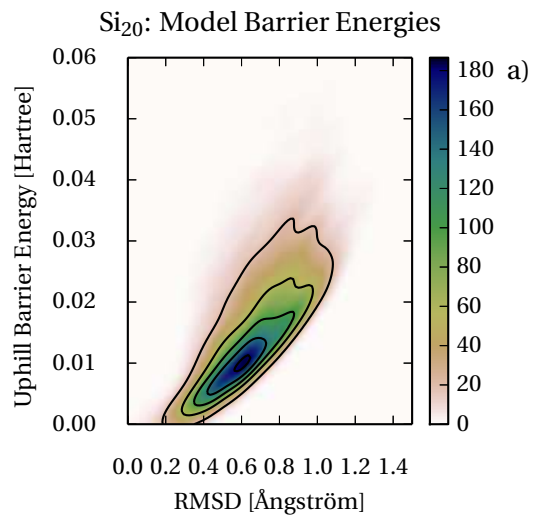}
  \caption{\label{fig:si20modelbarr} Same as
Fig.~\ref{fig:si20minenerdiff}
but for model uphill barrier energies instead of energy differences of
minima. Shown is the distribution of
uphill barriers plotted versus the configurational distance of
directly neighboring minima, as obtained by the model of
Eq.~\ref{eq:paraboluphillbarr}. Here, the same pairs of minima are used that
already were used for Fig.~\ref{fig:si20minenerdiff}.}
\end{figure}

In order to analyze this idea more rigorously, a simple parabola model
of the PES, as illustrated in Fig.~\ref{fig:rmsdbemodel}, is used.  In
fact, similar parabola models can be used for the explanation of the
Bell-Evans-Polanyi principle (a linear model is sufficient, though),
the Marcus equation, Hammond's postulate and the relationship of
low-curvature directions with low barrier
energies.\cite{Bell1936,Evans1936,Hammond1955,Marcus1968,Jensen2007,Roy2008,Sicher2011}
In such a
parabola model, the transition state is given by the intercept
$(\xi,E(\xi))$ of both parabolas. From Fig.~\ref{fig:rmsdbemodel} it is
evident that the barrier energies should rise with increasing structural distances
between the minima.
Here both parabolas are assumed to have the same curvature $k$
(``force constant''), and their minimum values are shifted by an amount
of $\epsilon$. The structural distance of both minima is denoted as
$a$. Consequently, the transition state $\xi$ and its corresponding
uphill barrier $E_{\text{u}}=E(\xi)$ is given by
\begin{align}
\xi &= \frac{a}{2} + \frac{\epsilon}{2ak},\\
E_{\text{u}} &= k\left(\frac{a}{2} +
\frac{\epsilon}{2ak}\right)^2.\label{eq:paraboluphillbarr}
\end{align}
For each pair of minima, this model is applied to the data of
Fig.~\ref{fig:si20minenerdiff} and the result is visualized in
Fig.~\ref{fig:si20modelbarr}a ($k=0.08\, \text{Ha}/\AA^2$). In contrast to the energy
differences of the minima in  Fig.~\ref{fig:si20minenerdiff}, this
model predicts a clear correlation between the
structural difference (RMSD) of two directly neighboring minima and the
energy of the corresponding uphill barrier.

It remains to be verified if the energies of real (computed) uphill
barriers between structurally very different minima also tend to be
larger than the energies of the uphill barriers between structurally
similar minima. If there is a breakdown in this hypothesis, it is
expected that no correlation of the type shown in
Fig.~\ref{fig:si20modelbarr} is seen.  For this verification,
transition states and their directly connected minima were computed for
$\text{Si}_{20}$ and $\text{Au}_{26}^{-}$ at the DFT level of theory as
implemented in the BigDFT\cite{Genovese2008,Mohr2014,Willand2013} code and for
$(\text{NaCl})_{32}$ and $(\text{NaCl})_{29}$ using the
Born-Mayer-Huggins-Tosi-Fumi\cite{Born1932,Mayer1933,Huggins1933,Fumi1964,Tosi1964}
(BMHTF) force field. For $\text{Si}_{20}$ the PBE\cite{Perdew1996}
functional was used, whereas for $\text{Au}_{26}^{-}$ the
LDA\cite{Kohn1965,ParrYang94} functional was used and in case of the
BMHTF force field the parameters of Ref.~\onlinecite{Adams1975} were
chosen.  Furthermore, transition states and the directly connected
neighbors were computed for the
Lennard-Jones\cite{Lennard-Jones1924,Lennard-Jones1925} clusters of
sizes 19, 38 and 55.  Except for $\text{Au}_{26}^{-}$, the geometries
and energies of the minima, as well as their connectivity, were
established using the MHGPS method as implemented in the BigDFT suite.
In the case of $\text{Au}_{26}^{-}$ the data was taken from a previous
study\cite{} and it is referred to this study for a description of its
computation.\cite{Schaefer2014a}  The density plots of the uphill
barrier energies versus the RMSD are given in
Fig.~\ref{fig:uphillbarrdist}. As can be seen from this figure, there
is indeed a good correlation between the structural difference (RMSD)
and the uphill barrier.
\begin{figure*}
\centering
  \begin{tabular}{@{}cc@{}}
    \includegraphics[scale=0.8]{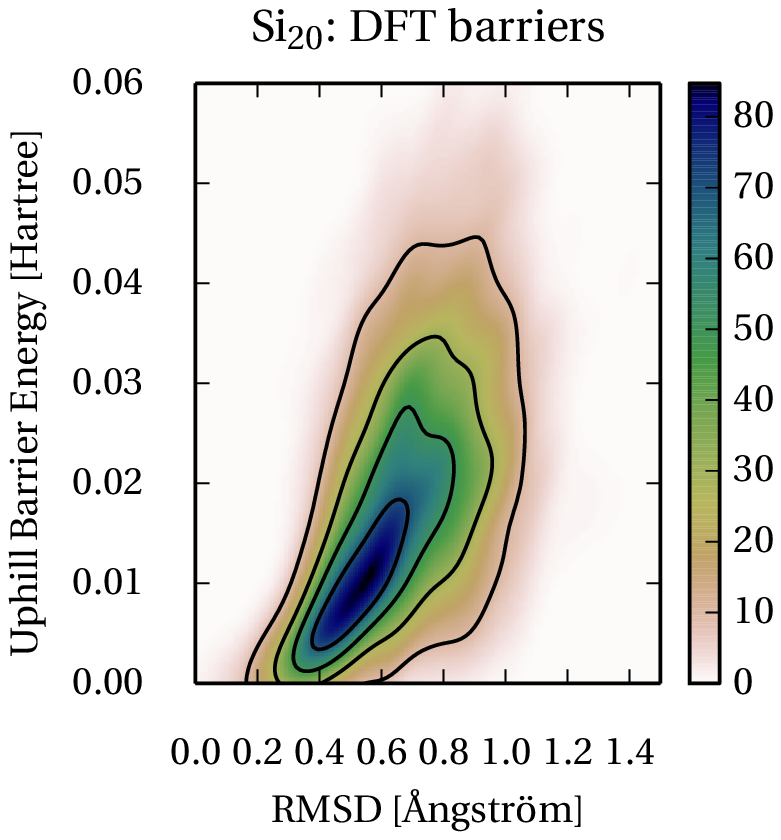} &
    \includegraphics[scale=0.8]{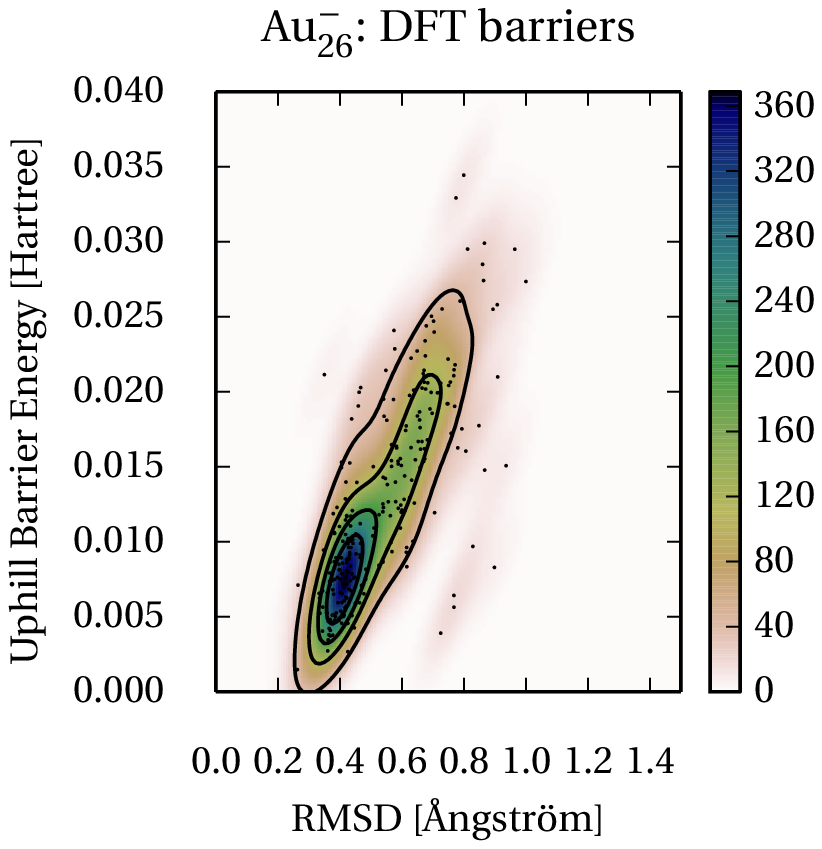}\\
    \includegraphics[scale=0.8]{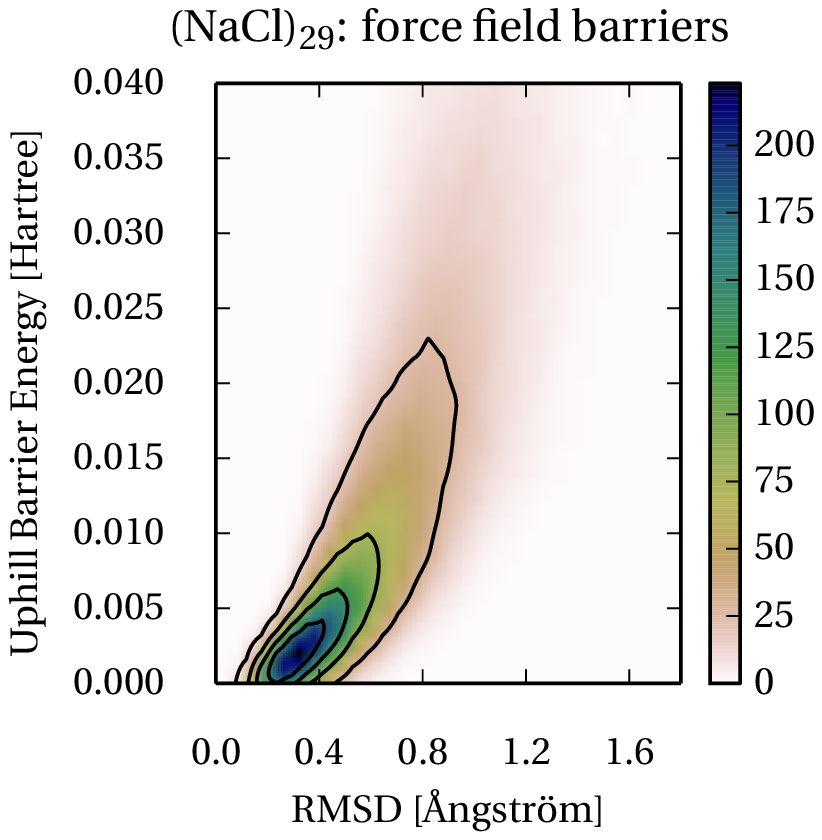}&
    \includegraphics[scale=0.8]{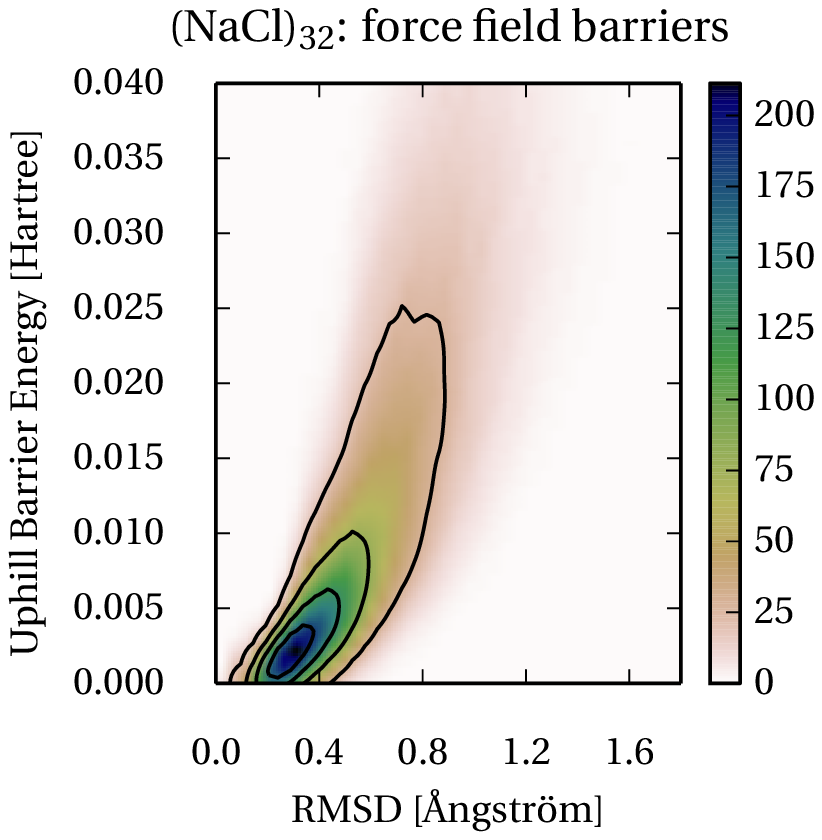}\\
  \end{tabular}
  \caption{\label{fig:uphillbarrdist} Gaussian kernel density estimates of the uphill barrier energies versus the
(permutationally and chirally optimized) RMSD distance of minima pairs
that are separated by only one intermediate transition state.  If two
minima are connected by more than one intermediate transition state,
only the transition state with the lowest energy was included in the
data sets used for these plots. The plot for $\text{Au}_{26}^{-}$ was obtained from only
259 transition states. It, therefore, is possible to show every single
data point for $\text{Au}_{26}^{-}$, which allows to demonstrate the soundness of the
Gaussian kernel density estimate.
The plot for $\text{Si}_{20}$ was generated from roughly 3,000
transition states and the plots for the systems described by force
fields were obtained from roughly 50,000 to 70,000 transition states.}
\end{figure*}
\renewcommand{\thefigure}{\arabic{figure}
(\textit{Continued}.)}
\begin{figure*}
\ContinuedFloat
\centering
  \begin{tabular}{@{}cc@{}}
    \includegraphics[scale=0.8]{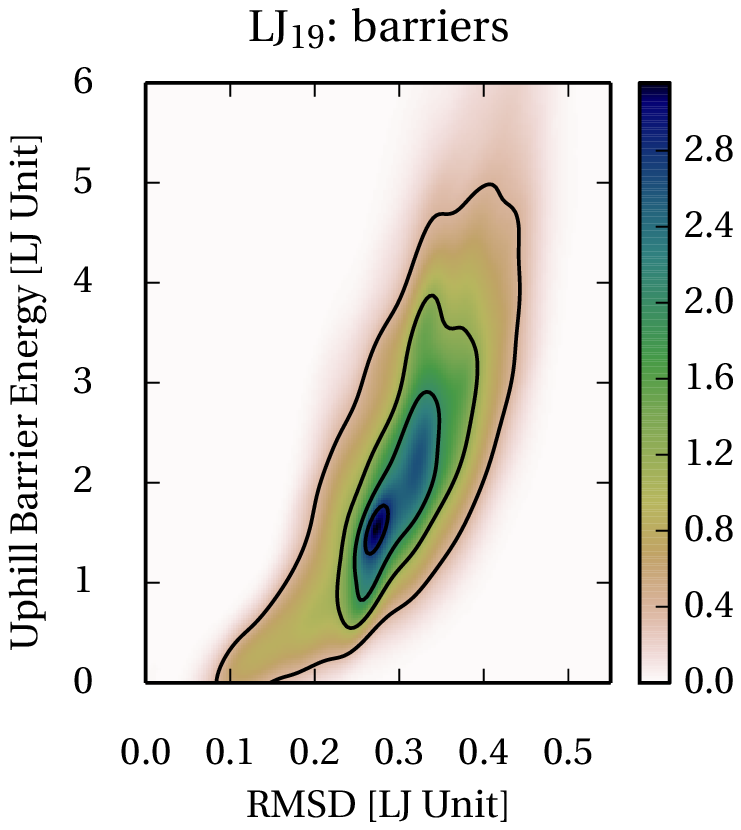}&
    \includegraphics[scale=0.8]{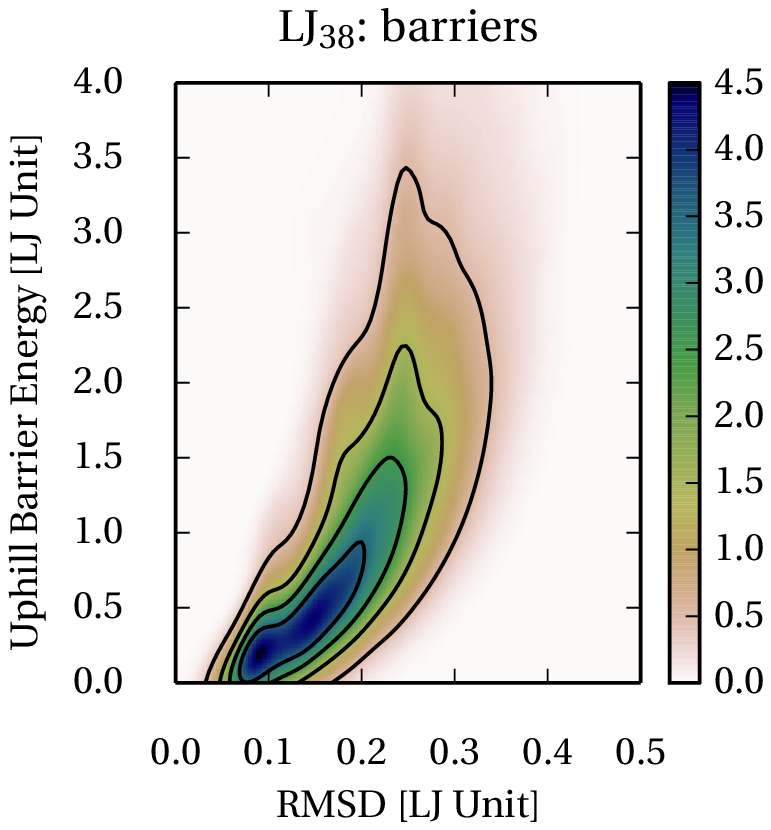}\\
\multicolumn{2}{c}{\includegraphics[scale=0.8]{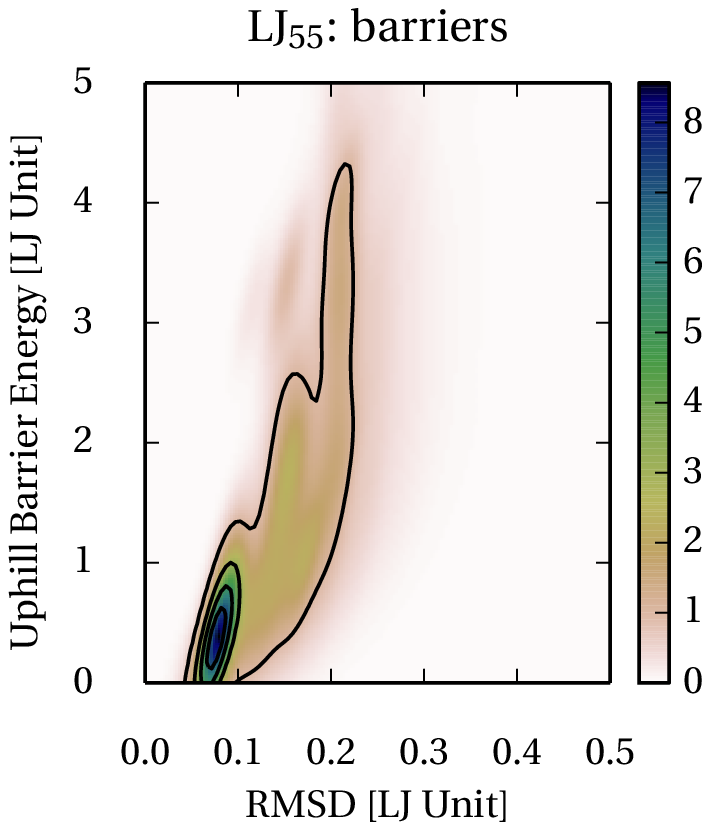}}
  \end{tabular}
  \caption{\label{fig:uphillbarrdistcont}}
\end{figure*}
\renewcommand{\thefigure}{\arabic{figure}}
Though the permutationally optimized RMSD is a very natural measure for
structural differences, it is very time consuming to compute, which
often makes it impracticable to use. For example, the computation of
the roughly 58,000 RMSDs for the $\text{LJ}_{55}$ plot in
Fig.~\ref{fig:uphillbarrdist} took about 14 hours (wall clock time),
despite using 150 cores in parallel. Of course, actual wall clock times
depend very strongly on the underlying computer hardware. Nevertheless, this
example illustrates that computing large numbers of RMSDs can be
problematic in practice.  Therefore, the plots of
Fig.~\ref{fig:uphillbarrdist} have been repeated using \textit{s}- and
\textit{p}-orbital fingerprint distances instead of RMSDs and are shown in
Fig.~\ref{fig:uphillbarrdistFP}. Again, a correlation between the
structural difference measured by the \textit{s}- and
\textit{p}-orbital fingerprint distance and the uphill barrier energy can be
observed.  Using \textit{s}- and \textit{p}-orbital based fingerprint
distances as a measure for
structural differences, the $\text{LJ}_{55}$ plot in
Fig.~\ref{fig:uphillbarrdistFP} took on the order of minutes on a
single core, which is a striking advantage over the RMSD and makes it
much more useful in practice. Plots from fingerprint distances using only
\textit{s}-type orbitals have a very similar appearance and are given
in the supplementary material.
\begin{figure*}
\centering
  \begin{tabular}{@{}cc@{}}
    \includegraphics[scale=0.8]{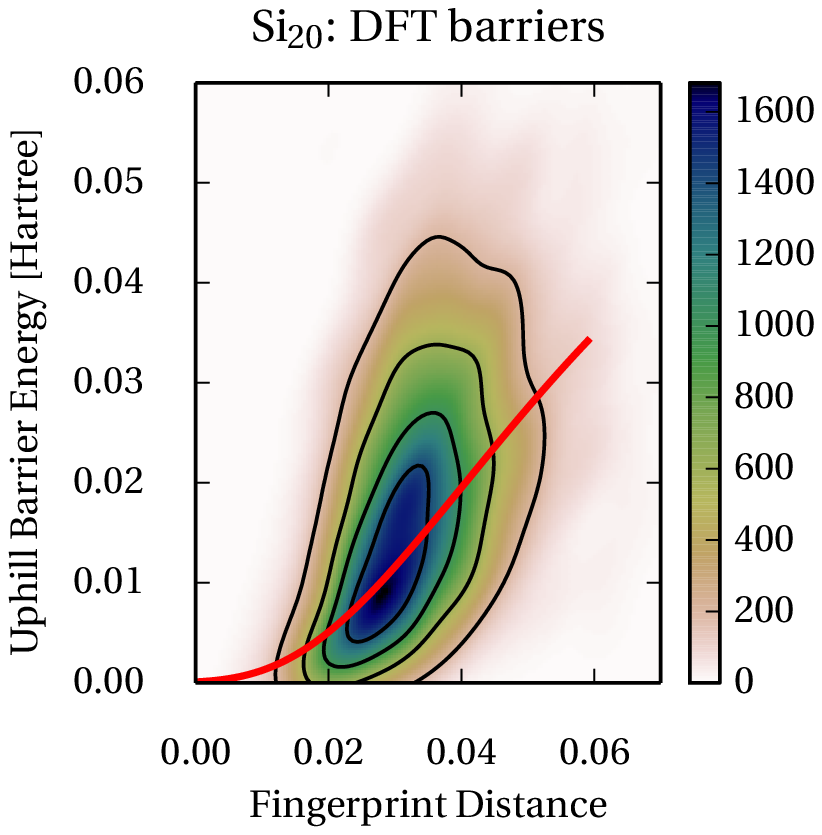}
&
    \includegraphics[scale=0.8]{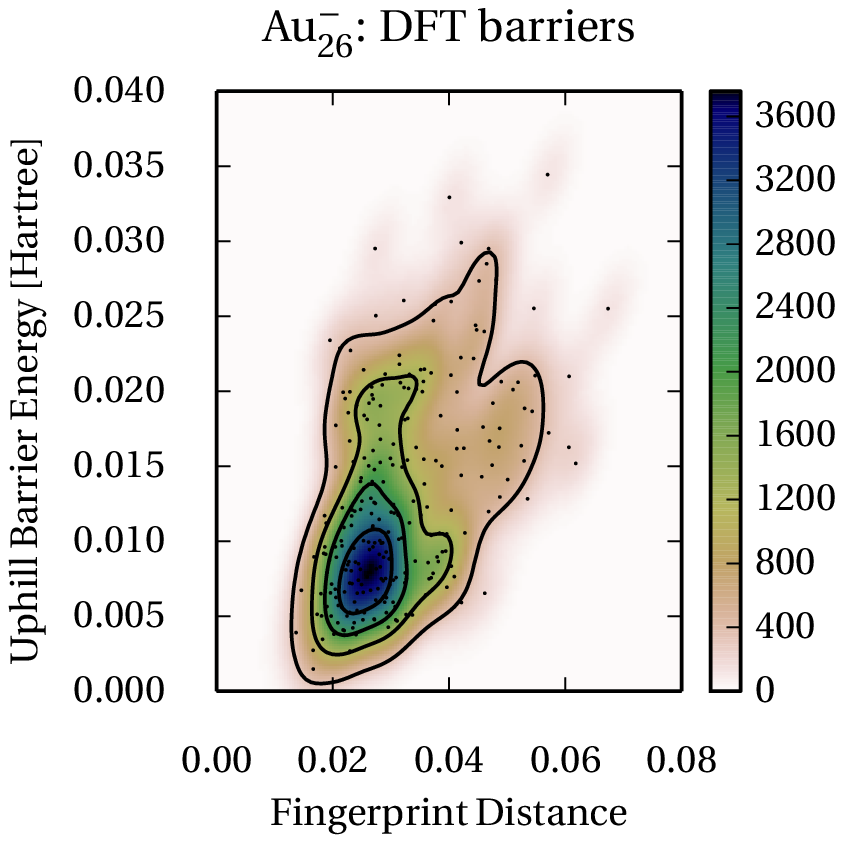}\\
    \includegraphics[scale=0.8]{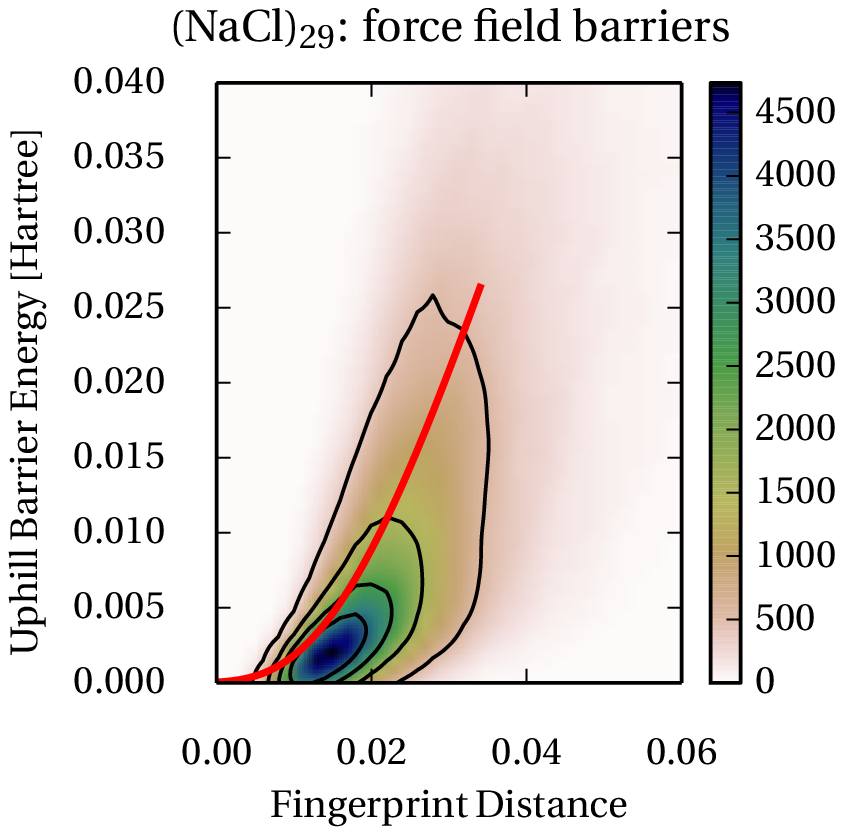}&
    \includegraphics[scale=0.8]{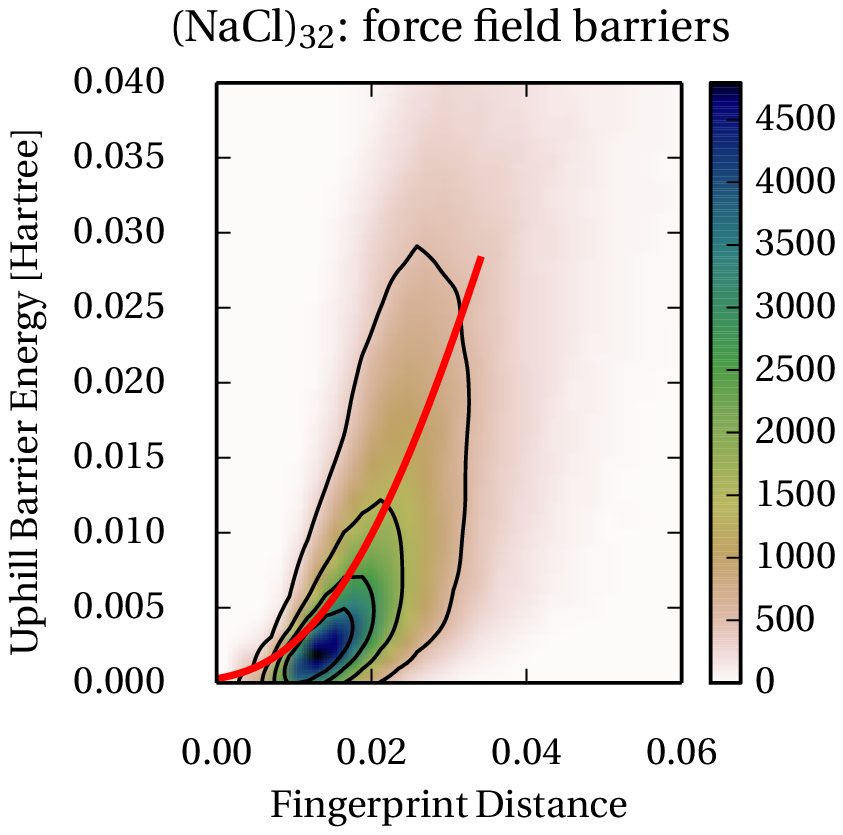}
  \end{tabular}
  \caption{\label{fig:uphillbarrdistFP} Same as
Fig.~\ref{fig:uphillbarrdist}, but using \textit{s}- and
\textit{p}-orbital fingerprint distances instead of the permutationally
optimized RMSD. Plots from fingerprint distances using only \textit{s}-type
orbitals have a very similar appearance and are given in the
supplementary material. The red lines are graphs of
Eq.~\ref{eq:fittingfct} and are discussed in Sec.~\ref{sec:fptrees}.}
\end{figure*}

\renewcommand{\thefigure}{\arabic{figure}
(\textit{Continued}.)}

\begin{figure*}
\ContinuedFloat
\centering
  \begin{tabular}{@{}cc@{}}
    \includegraphics[scale=0.8]{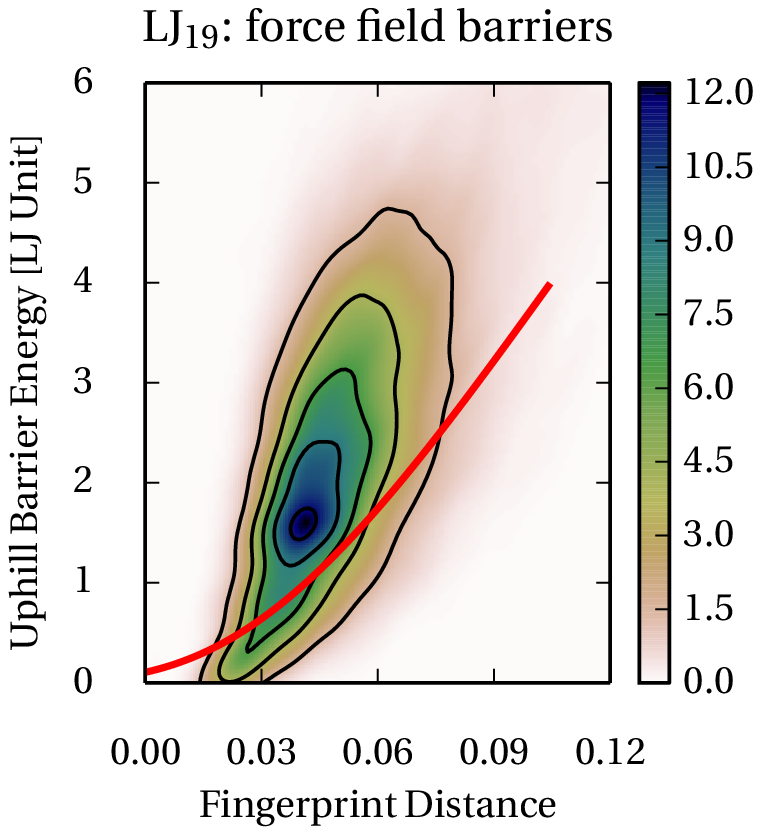}&
    \includegraphics[scale=0.8]{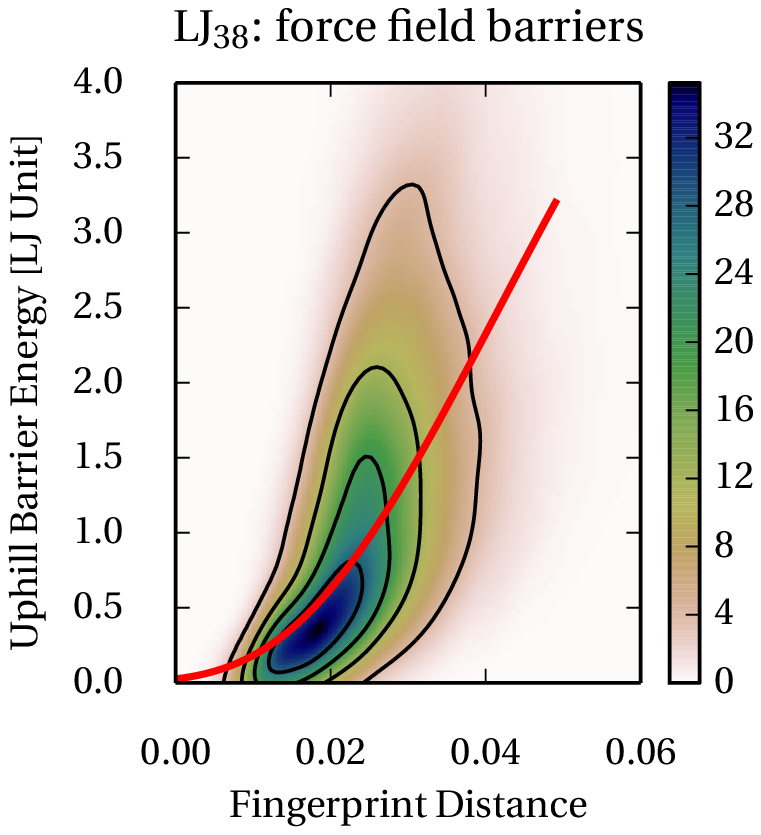}\\
\multicolumn{2}{c}{\includegraphics[scale=0.8]{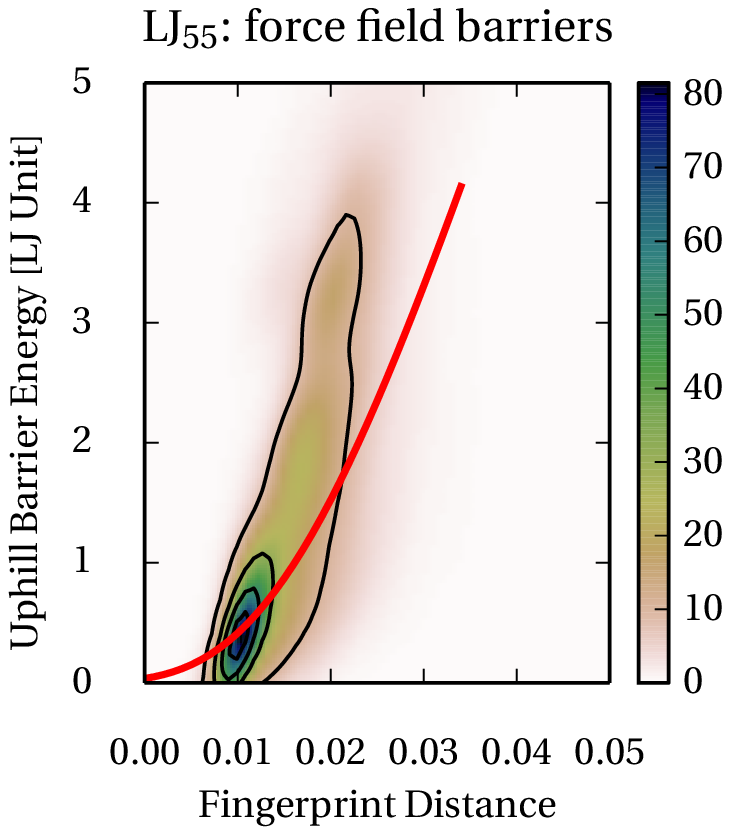}}
  \end{tabular}
  \caption{\label{fig:uphillbarrdistFPcont}}
\end{figure*}
\renewcommand{\thefigure}{\arabic{figure}}

Finally, a short comment seems to be appropriate on why it is almost
exclusively focused on the uphill barriers. After all, as can be seen
from Eq.~\ref{eq:paraboluphillbarr}, the same dependence of the
downhill barriers on the structural distance as for the uphill barriers
is predicted, except for a constant energy shift that is given by the
energy difference of both minima. This, however, does not imply that
necessarily a similar correlation as for the uphill barriers must be
observed for the downhill barriers. The reason is, that even though two
minima might be far apart from each other, the downhill barrier can be
vanishingly small if, in return, the energy difference between the two
minima is comparatively large. Indeed, plotting the downhill barrier
versus the structural difference results in a distribution that looks
very similar to the distribution of the energy differences of the
minima. 

\section{Generating Rough Overviews of Potential Energy
Surfaces}\label{sec:fptrees}

In this section, an empirical method suitable to generate trajectory-based
connectivity databases is presented. This method is based on post-processing data obtained
from one or several MH runs. Once MH runs are done, the computational
cost of this method is independent of the underlying
level of theory that was used for the MH runs. On a single core of a
standard office computer, this method allows the generation of trajectory-based
connectivity databases within a negligible amount of wall clock time,
even if the trajectory-based connectivity database shall describe PESs that
are defined by computationally demanding methods, like for example DFT.
To introduce this novel method, first the term ``trajectory-based
connectivity database'' is defined. A trajectory-based
connectivity database
is understood to contain three types of information. First, it contains
all local minima visited during a certain number of MH runs. Second, it
contains the information which minima were visited consecutively by the
MH walkers and finally, also a qualitative measure for the energy
needed to interconvert the consecutively visited minima is part of a
trajectory-based
connectivity database. Furthermore, a pair of minima
visited consecutively by the MH walker will be denoted as ``hopping
pair''.

In contrast to such a trajectory-based connectivity database, the stationary
point database defined by Wales\cite{Wales2002,Wales2003,Wales2004-2}
contain minima, transition states and the information to which minima
the transition states are connected by minimum energy or energy
minimized pathways.  Thus, a trajectory-based connectivity database can be
seen as an approximation to a stationary point database. The
connectivity information is approximated by the information which
minima were visited consecutively by the MH walker. This is a
reasonable approximation, because the MH walkers explore the PES by
means of short MD trajectories that, at most times, have relatively moderate
initial kinetic energies. As a consequence, the geometries of
hopping pair members typically are very similar to each other, a
circumstance that is also used in the MHGPS scheme.\cite{Schaefer2014}
\begin{figure*}
\centering
\includegraphics{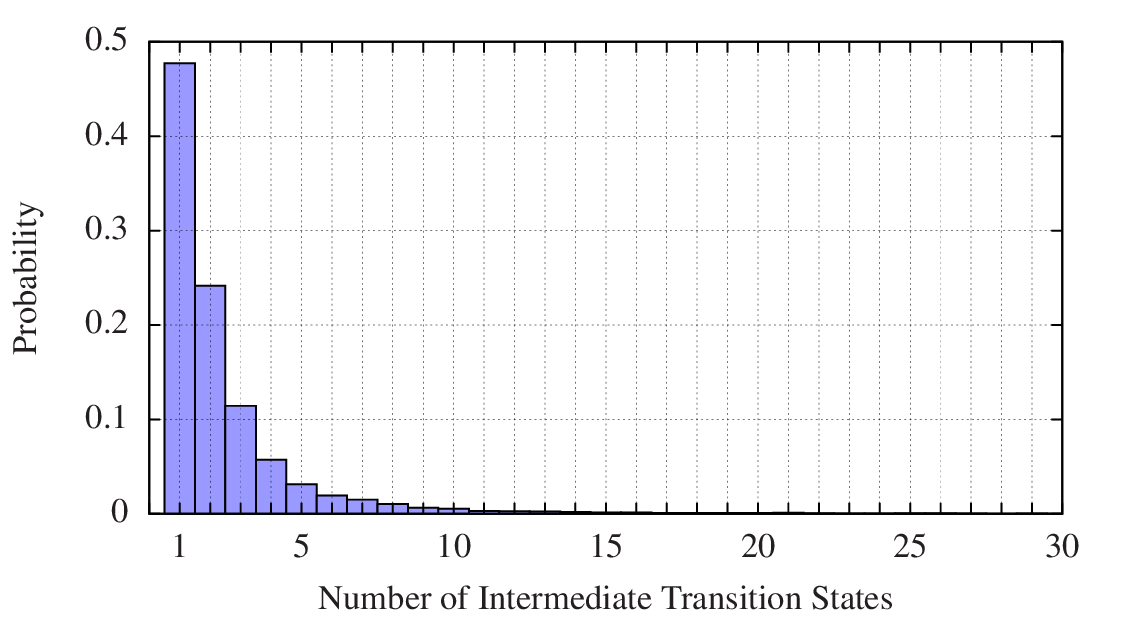}
\caption{\label{fig:histolj55} Shown for the $\text{LJ}_{55}$ system is
the probability distribution of the number of intermediate transition states
needed by the MHGPS approach as implemented in the BigDFT-suite to
connect pairs of consecutively accepted minima.  The data set consists
of more than 20,000 connection attempts that were stopped if the
connection could not be established within 30 transition state
computations.  }
\end{figure*}
Quantitative evidence for the validity of this
connectivity approximation is given in Fig.~\ref{fig:histolj55}. In this
figure, the probability distribution of the number of intermediate transition
states needed by the MHGPS method to connect pairs of consecutively
accepted minima is given. These numbers constitute an upper bound to
the minimum number of intermediate transition states located in between
two consecutively accepted minima. It can be seen from this figure that
the majority of consecutively accepted minima can be connected with
each other by no more than two intermediate transition states.

What remains to be discussed is, how an educated guess for the energy,
which is needed to interconvert the minima of a hopping pair, can be
obtained. Before describing the actual method for obtaining such a
guess, a different approach is discussed. From a theoretical point of
view, it would be very satisfying if Eq.~\ref{eq:paraboluphillbarr}
could be used to obtain a guess for the transition state energy.
Indeed, using a suitable value for the force constant $k$, it turned
out to be possible to generate disconnectivity graphs of similar
quality as those based on the method that is presented below.  However,
for us, it was only possible to choose good values for $k$, if the
correct appearance of the disconnectivity graph was known.
Unfortunately, a procedure that is able to reliably determine the force
constant and that is able to give disconnectivity graphs of similar
quality as those based on the method outlined below has yet to be
found. In fact, using inappropriate values for $k$ can produce
completely misleading disconnectivity graphs. In contrast to this, in
all tested cases, the approach discussed below produced qualitatively
very reasonable disconnectivity graphs.

\begin{figure*}
\centering
\includegraphics{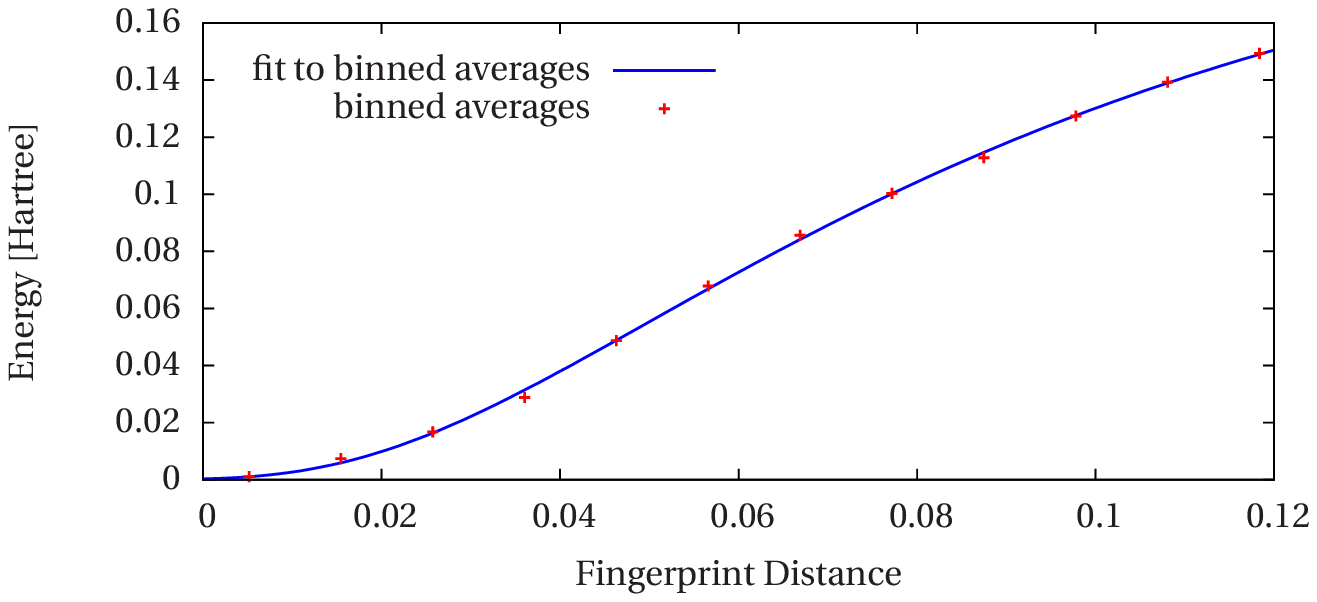}
\caption{\label{fig:fit}
Fit of $E_{\text{u}}$ as defined in Eq.~\ref{eq:fittingfct} to the
binned averages of the energy differences of $(\text{NaCl})_{32}$
hopping pairs, as modeled by the BMHTF force field, versus
their structural difference measured by the overlap matrix fingerprint
distance using \textit{s}- and \textit{p}-type orbitals. 25
bins were used for grouping the roughly 28.000 data points. Of those 25
bins, only those that contain at least 5\% of the data points
of the bin with the most data points are shown and were used for the
fit. The values of the fitting parameters are $\alpha=0.2449$ Ha,
$\beta=0.0128$, $\gamma=0.0445$ and $\delta=-2.0159$.
}
\end{figure*}

The remainder of this section focuses on describing the empirical
method that, was able to produce an educated qualitative guess for the
transition state energies.  In this approach the energy difference of
the two minima of a hopping pair is compared to the average energy
difference of minima of hopping pairs that are separated by a
similar structural fingerprint distance. If the energy difference is
larger than the average value at this fingerprint distance, the uphill
barrier of a hopping pair is estimated as the absolute value of the
energy difference of the two hopping pair members. Otherwise, the
uphill barrier is estimated as the average absolute value of the energy
differences at this fingerprint distance.  In practice, this is done by
plotting the absolute values of the energy differences of the minima of
each hopping pair versus their fingerprint distance and  computing
binned averages of this data. A continuous function describing this
binned average is obtained by means of a fitting procedure.  Of course,
this approach does not give a quantitative estimate of the energy of
each single barrier, but it is intended to reproduce the energy scale
and roughly the average trend in uphill barrier energies that was
discussed in the previous section.  More explicitly, assuming the
minima energies of a hopping pair to be $E_1$ and $E_2$ with $E_1
\le E_2$, the absolute energy $E_{\text{t}}$ needed to interconvert the
two minima is estimated as
\begin{align}
E_{\text{t}} :=
\text{max}\left(E_1+E_{\text{u}}(a),\,E_2\right),\label{eq:barrfct}
\end{align}
where the max-function returns the larger of its two arguments and the
uphill barrier energy $E_{\text{u}}$ is a function of the
fingerprint distance $a$ (see Fig.~\ref{fig:rmsdbemodel}). $E_{\text{u}}$ is defined as
\begin{align}
E_{\text{u}}(a) := \alpha\exp(-\beta\lvert a
+\gamma\rvert^\delta),\label{eq:fittingfct}
\end{align}
where the parameters $\alpha,\,\beta,\,\gamma$ and $\delta$ are
obtained by a fit to the binned averages of the energy differences of
the minima of hopping pairs. The fitting function given in
Eq.~\ref{eq:fittingfct} is a heuristic and pragmatic choice that turned
out to work well in all tested cases.
 The fitting
itself is performed with the help of the nonlinear least-squares
Marquardt-Lavenders algorithm as implemented in the gnuplot
code.\cite{Levenberg1944,Marquardt1963,gnuplot} Of course, other
fitting methods can be used, because $E_{\text{u}}$ is only required to
provide a continuous function of the qualitative trends for the uphill
barrier energies. A plot exemplifying such a fit is given  in
Fig.~\ref{fig:fit} for the case of $(\text{NaCl})_{32}$.

It turned out that by using Eq.~\ref{eq:barrfct} for obtaining transition
state energy guesses, it is possible to produce disconnectivity graphs that
reflect reasonably well the characteristics of a PES.
Before presenting these disconnectivity graphs, it is appropriate to  discuss the
reasonable performance of Eq.~\ref{eq:barrfct}. To see this, first it is realized that Eq.~\ref{eq:barrfct}
splits up the hopping pairs into two sets.

In the first set, the uphill barrier of a hopping pair is guessed by
means of Eq.~\ref{eq:fittingfct}.  In Fig.~\ref{fig:uphillbarrdistFP},
the fitting function Eq.~\ref{eq:fittingfct} is plotted on top of the
uphill barrier distributions of $\text{Si}_{20}$, $(\text{NaCl})_{29}$,
$(\text{NaCl})_{23}$, $\text{LJ}_{19}$, $\text{LJ}_{38}$ and
$\text{LJ}_{55}$. From these plots it is evident that the binned
average of the absolute values of the energy differences of hopping
pair minima is a reasonable guess for the uphill barrier energy.
Eq.~\ref{eq:fittingfct} prevents the assignment of low transition state
energies to hopping pairs with structurally very different minima and,
therefore, is in agreement with the results of
Sec.~\ref{sec:corrTSener}. This agreement is essential for an
acceptable reproduction of the characteristics of a PES. Otherwise, as
will be seen from the disconnectivity graphs that are presented below,
superbasins are likely to be merged, which can result into a completely
misleading appearance of a PES. Furthermore it can be seen from
Fig.~\ref{fig:uphillbarrdistFP} that the uphill barrier energy which is
assigned to a hopping pair corresponds in most cases to a not
completely unlikely uphill barrier energy at a given structural
distance.  As was demonstrated by Fig.~\ref{fig:histolj55}, the minima
of many hopping pairs are separated by only one intermediate transition
state and it is clear that the trend of increasing uphill barrier
energies with increasing structural distances that was described in
Sec.~\ref{sec:corrTSener} can be applied to these hopping pairs.
However, there is no strict guarantee for the minima of a hopping pair
to be in a close neighborhood to each other.  Despite this fact, it is
still the trend that was described in
Sec.~\ref{sec:corrTSener} that is used to obtain a guess for the barrier
energies of those hopping pairs. At a first glance, this might be
surprising since two structurally very different minima, which only can
be interconverted into each other by crossing many intermediate
transition states, might very well be separated by a low overall
barrier. For example, this can be the case if the pairwise structural
distances of all intermediate minima are small. Using a measure for the
transition state energies that is based on the correlation discussed in
Sec.~\ref{sec:corrTSener}, a high barrier energy will be assigned to
the direct transition between such minima.  However, this is not a
disadvantage, but rather a desirable effect.  Typically, the analysis
of a trajectory-based connectivity database will focus on low energy
pathways.  In such an analysis, the direct interconversion of those far
apart minima is disfavored due to the high energy that is assigned to
their direct interconversion. In contrast, low barrier energies are
properly assigned to the pathway that leads over the large number of
pairwise structurally similar minima, which
allows for its identification.

In the second set, the uphill barriers of hopping pairs are approximated by the energy of
the energetically
higher minimum. For transitions with downhill
barriers that are small compared to the uphill barrier, this is
a sufficient approximation. However, if the energy difference between two
minima is small and their structural difference large, this approximation
is not only quantitatively, but also qualitatively very inaccurate.
Fortunately, Eq.~\ref{eq:fittingfct} rigorously prevents the latter hopping pairs from being included into this second set. This second set only contains hopping pairs
with above-average energy differences with respect to a given
structural distance.
Therefore, for those hopping pairs for which a significant underestimation of
the transition state energy endangers a reasonable
reproduction of the overall PES characteristics in a
disconnectivity graph, the uphill barriers are not estimated by the energy difference of
the involved minima.

\begin{figure*}
\centering
  \begin{tabular}{@{}ccc@{}}
    \includegraphics[scale=0.9]{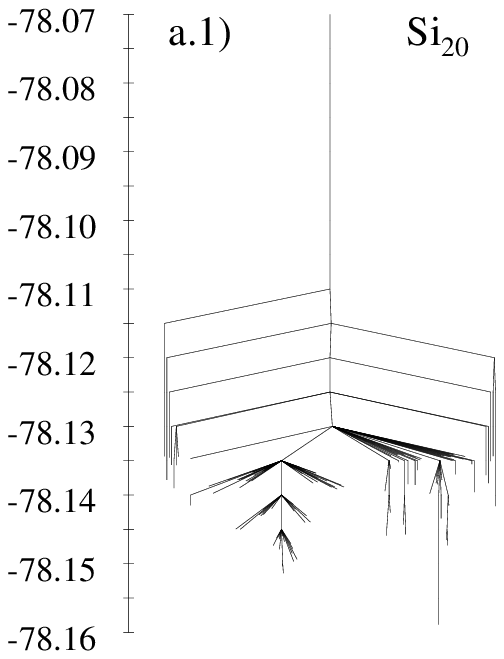} &
    \includegraphics[scale=0.9]{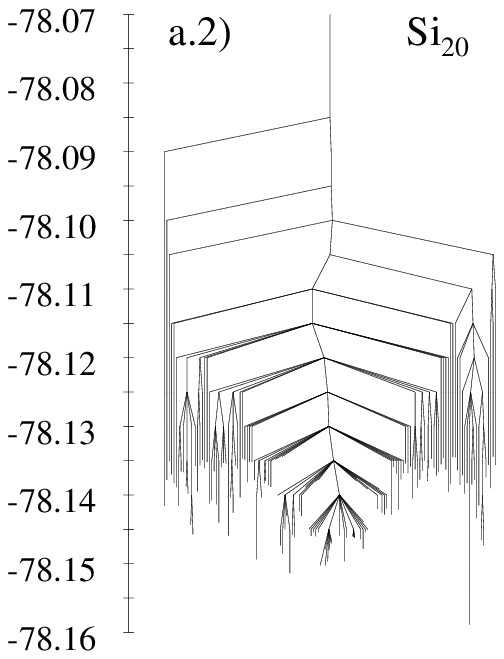} &
    \includegraphics[scale=0.9]{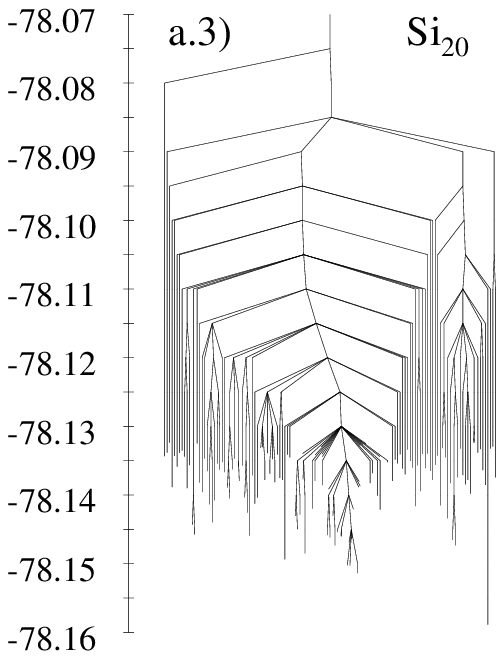} \\
    \includegraphics[scale=0.9]{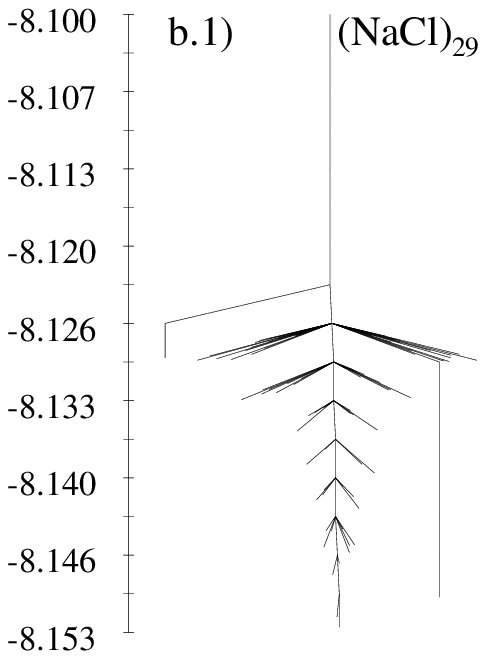} &
    \includegraphics[scale=0.9]{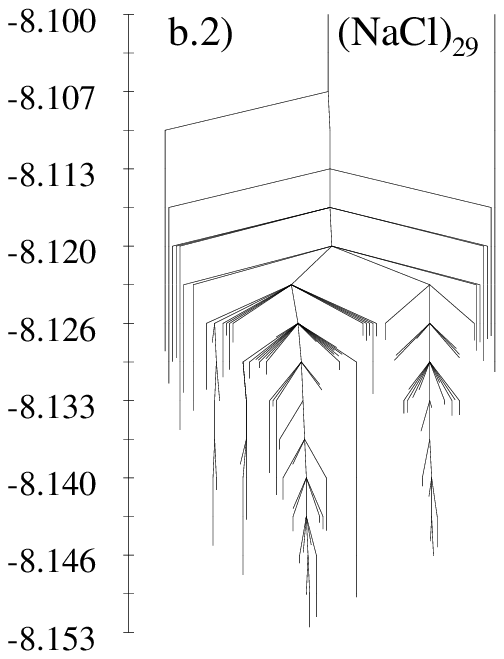} &
    \includegraphics[scale=0.9]{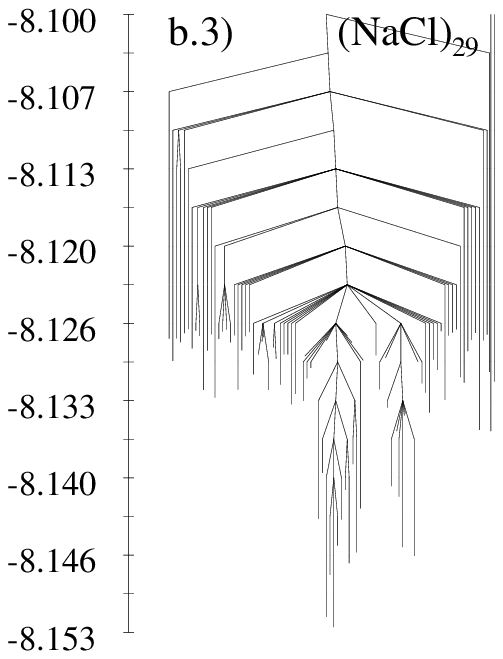} \\
    \includegraphics[scale=0.9]{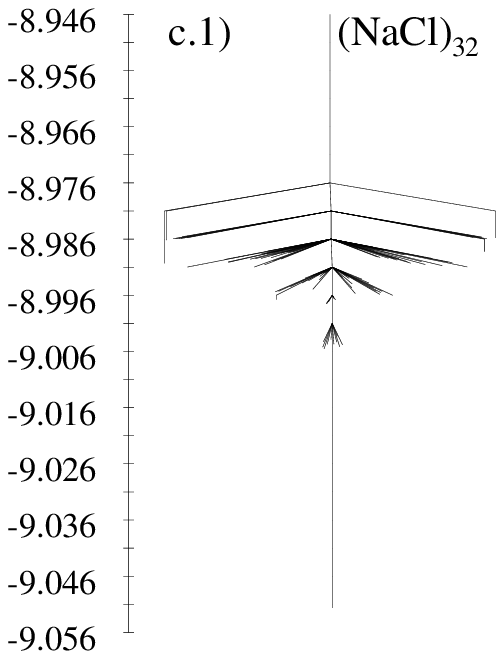} &
    \includegraphics[scale=0.9]{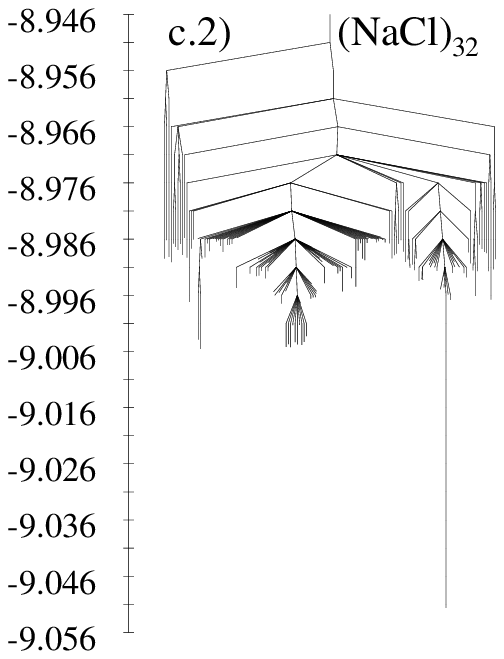} &
    \includegraphics[scale=0.9]{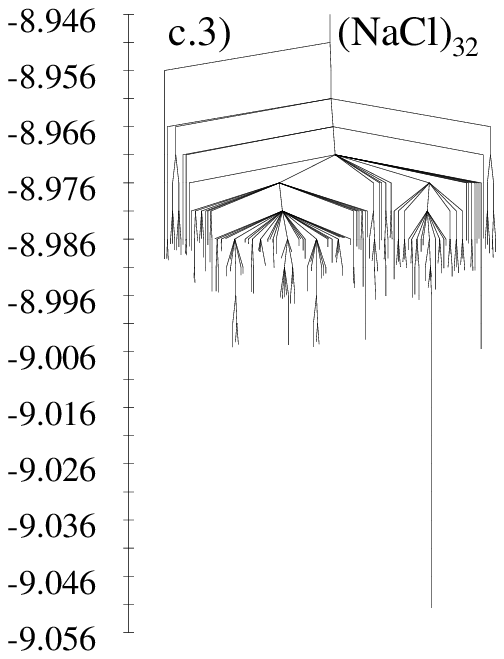} \\
  \end{tabular}
  \caption{\label{fig:fptrees} Disconnectivity graphs for
$\text{Si}_{20}$ (panels~(a.$n$)),  $(\text{NaCl})_{29}$ (panels~(b.$n$)),
$(\text{NaCl})_{32}$ (panels (c.$n$)),
$\text{LJ}_{19}$ (panels~(d.$n$)), $\text{LJ}_{38}$ (panels~(e.$n$))
and $\text{LJ}_{55}$ (panels~(f.$n$)). The graphs in panels~(x.1) and
(x.2) are based on trajectory-based connectivity databases. For the (x.1) panels, the barriers were eliminated, whereas
the approximations to the
barrier energies described in Sec.~\ref{sec:fptrees} were used for
the (x.2) panels. Reference graphs based on stationary point databases
that were sampled by the MHGPS approach are
shown in the rightmost column (panels~(x.3)). The energy scale is in
Hartree ($\text{Si}_{20}$, $(\text{NaCl})_{29}$, $(\text{NaCl})_{32}$) and
in Lennard-Jones units ($\text{LJ}_{19}$, $\text{LJ}_{38}$,
$\text{LJ}_{55}$). }
\end{figure*}
\renewcommand{\thefigure}{\arabic{figure}
(\textit{Continued}.)}
\begin{figure*}
\ContinuedFloat
\centering
  \begin{tabular}{@{}ccc@{}}
    \includegraphics[scale=0.9]{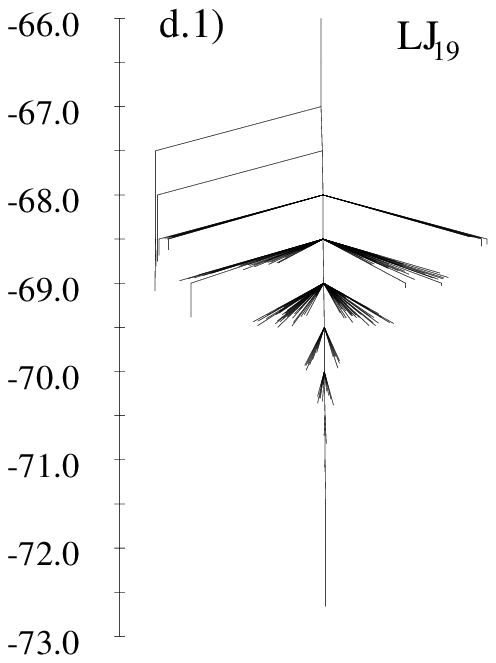} &
    \includegraphics[scale=0.9]{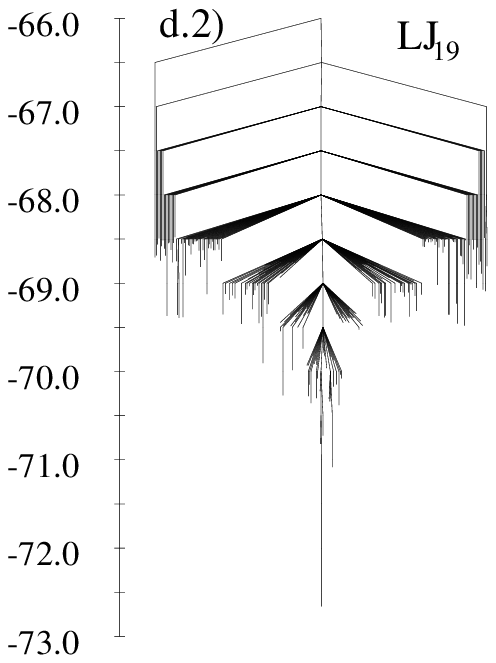} &
    \includegraphics[scale=0.9]{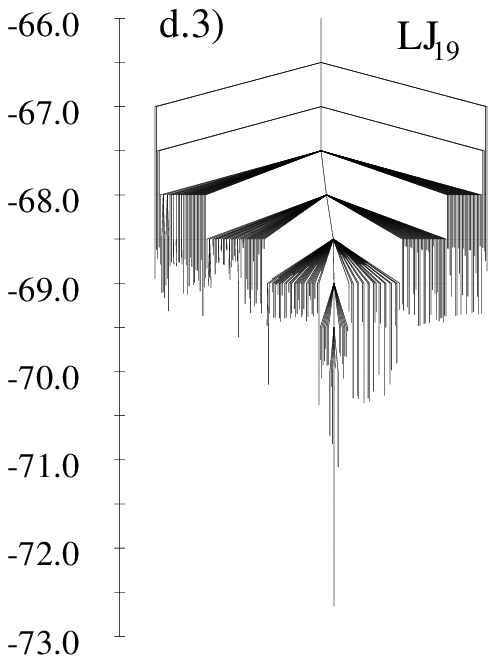} \\
    \includegraphics[scale=0.9]{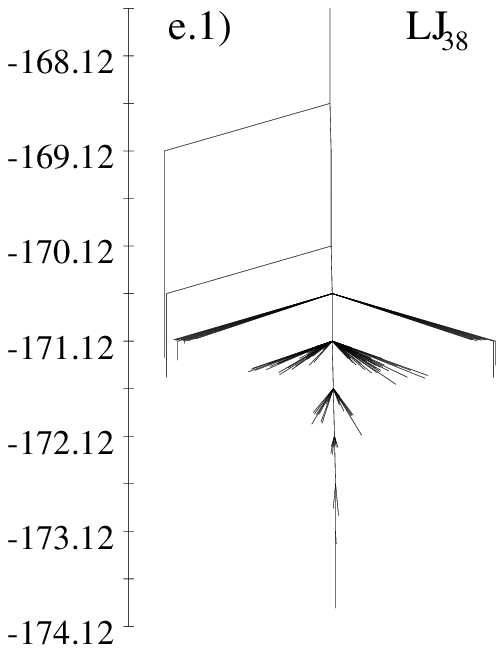} &
    \includegraphics[scale=0.9]{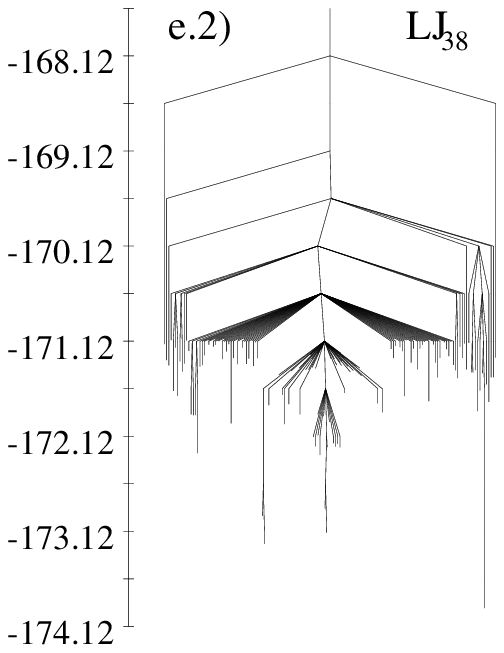} &
    \includegraphics[scale=0.9]{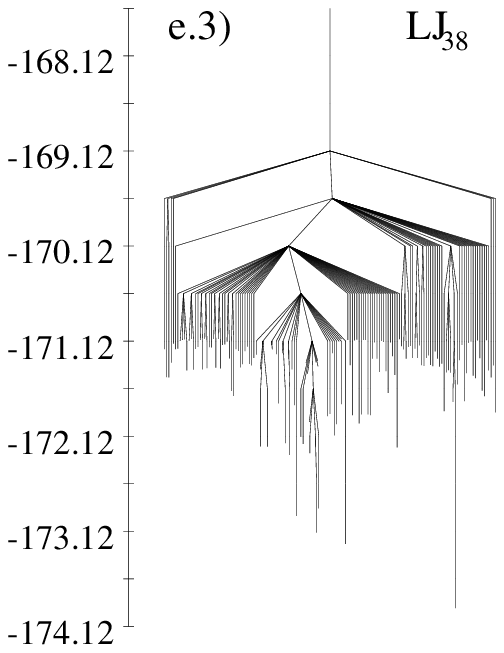} \\
    \includegraphics[scale=0.9]{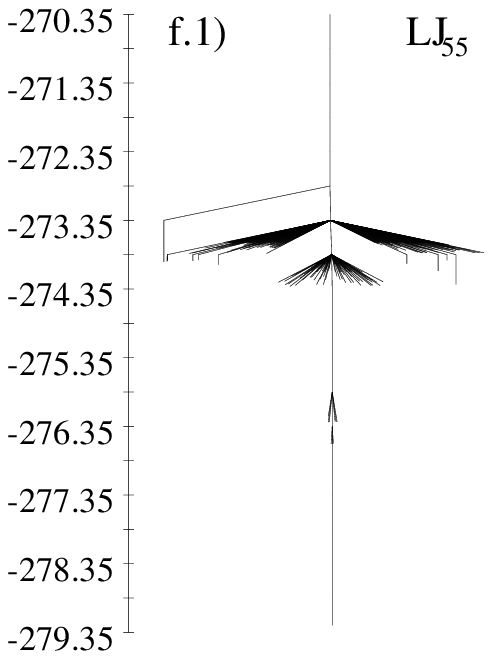} &
    \includegraphics[scale=0.9]{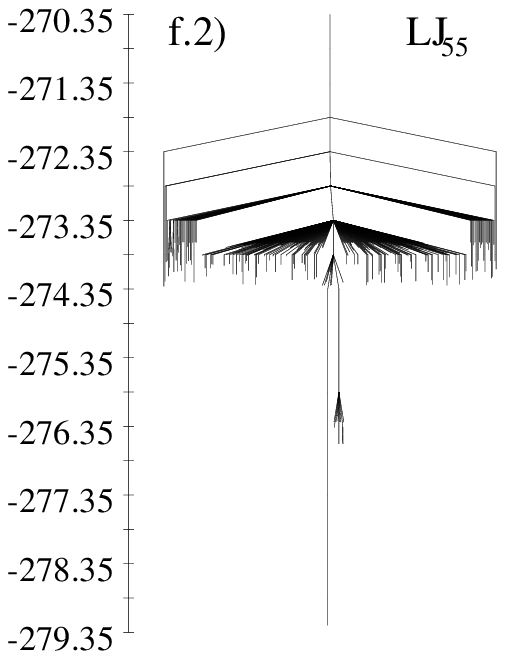} &
    \includegraphics[scale=0.9]{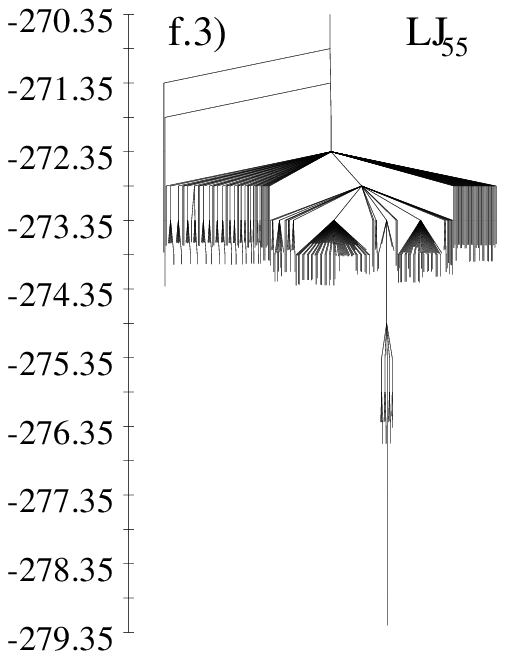}
  \end{tabular}
  \caption{\label{fig:fptreescont} }
\end{figure*}
\renewcommand{\thefigure}{\arabic{figure}}
Fig.~\ref{fig:fptrees} displays disconnectivity graphs for
$\text{Si}_{20}$, $(\text{NaCl})_{29}$, $(\text{NaCl})_{32}$,
$\text{LJ}_{19}$, $\text{LJ}_{38}$ and $\text{LJ}_{55}$. As above, the
PES of $\text{Si}_{20}$ was computed at the DFT level of
theory as implemented in the BigDFT code (PBE exchange correlation functional). For the sodium chloride
clusters, again the BMHTF force field was used.  No disconnectivity
graphs are presented for
$\text{Au}_{26}^{-}$ because only the local minima, but not the complete
minima hopping history, were archived from the previous
work.\cite{Schaefer2014a}
The panel labels of Fig.~\ref{fig:fptrees} follow the scheme (x.$n$),
where ``x'' is one of
a, b, c, d, e or f and represents the system (a=$\text{Si}_{20}$,
b=$(\text{NaCl})_{29}$, c=$(\text{NaCl})_{32}$, d=$\text{LJ}_{19}$,
e=$\text{LJ}_{38}$ and f=$\text{LJ}_{55}$) and $n$ runs from one to
three. Disconnectivity graphs in the panels $(x.1)$ and $(x.2)$ (the left and
middle column of Fig.~\ref{fig:fptrees}) are based on trajectory-based connectivity databases, where for the $(x.1)$ panels the barrier energies
were set to the energy of the higher minimum and for the $(x.2)$ panels
the barrier energies were approximated by Eq.~\ref{eq:barrfct} and
the above described fitting procedure. The (x.2) disconnectivity
graphs will also be denoted as ``fingerprint disconnectivity graphs''.
For the center column of Fig.~\ref{fig:fptrees}, fingerprint distances
based on \textit{s}-
and \textit{p}-orbitals were used.
Disconnectivity graphs in the
rightmost column of Fig.~\ref{fig:fptrees} (panels~(x.3)) are based on
stationary point databases that were generated by means of the MHGPS
approach.\cite{Schaefer2014} These standard disconnectivity
graphs are considered as the reference for the present purpose.
For each system, all three disconnectivity graphs show the same number
of minima, however, not necessarily the identical minima. This is,
because the stationary point databases are usually much more detailed,
because they were thoroughly sampled by the MHGPS
approach in order to generate exact reference disconnectivity graphs.
In Tab.~\ref{tab:dbsize} rough sizes of the underlying databases are given.
\begin{table}
\caption{Rough sizes of the databases used for Fig.~\ref{fig:fptrees}.}
\begin{ruledtabular}
\begin{tabular}{cccc}
database type  & system & $n$\footnotemark[1]  & $m$\footnotemark[2] \\
\hline
TBCD\footnotemark[3]  & $\text{Si}_{20}$      &  7,000       & 5,000  \\
                                       & $\text{(NaCl)}_{29}$  &  82,000      & 71,000  \\
                                       & $\text{(NaCl)}_{32}$  &  28,000      & 25,000  \\
                                       & $\text{LJ}_{19}$      &  1,800       & 1,100  \\
                                       & $\text{LJ}_{38}$      &  87,000      & 64,000  \\
                                       & $\text{LJ}_{55}$      &  35,000      & 33,000 \\
SPD\footnotemark[4]  &              $\text{Si}_{20}$      &  3,400       & 2,000  \\
                                       & $\text{(NaCl)}_{29}$  &  200,000     & 171,000  \\
                                       & $\text{(NaCl)}_{32}$  &  68,000      & 61,000  \\
                                       & $\text{LJ}_{19}$      &  65,000      & 14,000  \\
                                       & $\text{LJ}_{38}$      &  68,000      & 45,000  \\
                                       & $\text{LJ}_{55}$      &  59,000      & 49,000  \\
\end{tabular}
\end{ruledtabular}
\footnotetext[1]{Number of minima.}
\footnotetext[2]{Number of hopping pairs in case of trajectory-based connectivity databases or number of transition states in case of stationary point databases.}
\footnotetext[3]{trajectory-based connectivity database}
\footnotetext[4]{stationary point database}
\label{tab:dbsize}
\end{table}

\begin{figure*}[t]
\centering
\includegraphics{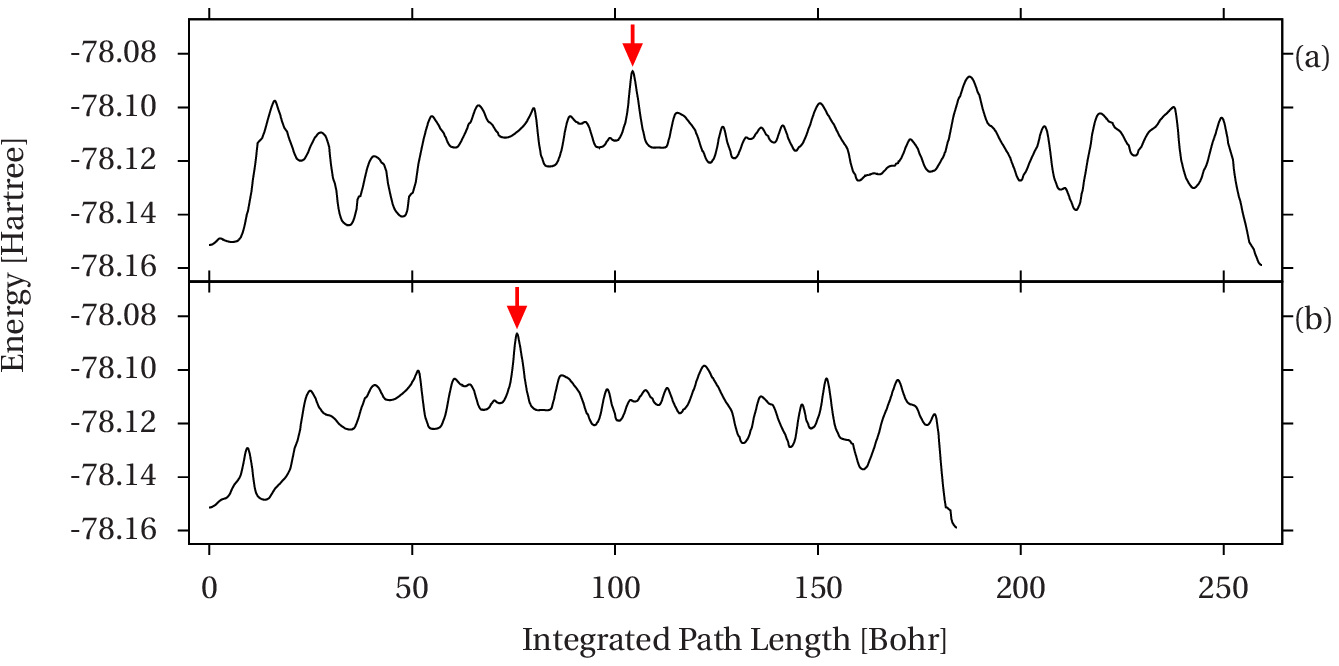}
  \caption{\label{fig:si20fpinputpath} Two energy minimized pathways connecting the two
lowest minima of $\text{Si}_{20}$ (DFT, PBE).
The pathway in panel~(a) was obtained by extracting a sequence of minima from the
trajectory-based connectivity database and using this sequence of minima as
input for the connectivity finder module of the MHGPS\cite{Schaefer2014} code. Panel~(b) shows a pathway that was extracted
from a stationary point database sampled by entirely unbiased MHGPS
runs.
The shown pathways are SQNM\cite{Schaefer2015} trajectories obtained by
relaxations from the
transition states after stepping away a small distance in both directions of the negative eigenmode. The
transition states in the MHGPS runs were tightly converged by means of
the SQNS\cite{Schaefer2015} method. The red arrows indicate the
highest energy transition states along the pathways. In both pathways,
the highest energy transition states are identical.
}
\end{figure*}

Even if only using the connectivity as provided by the trajectory-based connectivity database, but eliminating all barriers, the double-funnel
landscape of $\text{Si}_{20}$ is clearly visible
(Fig.~\ref{fig:fptrees}a.1),
nevertheless, the
appearance of the disconnectivity graph is improved by using the
fitting procedure for approximating transition state energies
(Fig.~\ref{fig:fptrees}a.2).
Though, for $\text{Si}_{20}$, the most important feature of the system is
already visible in the (a.1)~panel, the same is not true for the
remaining systems. Except for $\text{Si}_{20}$, completely eliminating
the barriers results in disconnectivity graphs that
correspond to extreme structure seekers and the true topology of the
PESs is not visible in the (x.1)~panels. In contrast to this, the
fingerprint disconnectivity graphs in the (x.2)~panels exhibit a
remarkable resemblance to the standard disconnectivity graphs shown in the
(x.3)~panels of Fig.~\ref{fig:fptrees}.

The fingerprint disconnectivity graphs based on \textit{s}- and
\textit{p}-orbital
fingerprints are slightly
 more similar to the standard disconnectivity graphs than those based only
on \textit{s}-orbitals and shown in the supplementary material. Nevertheless,
also the fingerprint disconnectivity graphs based on the \textit{s}-only
fingerprints provide
a striking resemblance to the standard disconnectivity graphs, in
particular if taken into account that generating fingerprint based disconnectivity
graphs is a quasi-free lunch post-processing of MH data.

\begin{figure*}
\centering
\includegraphics{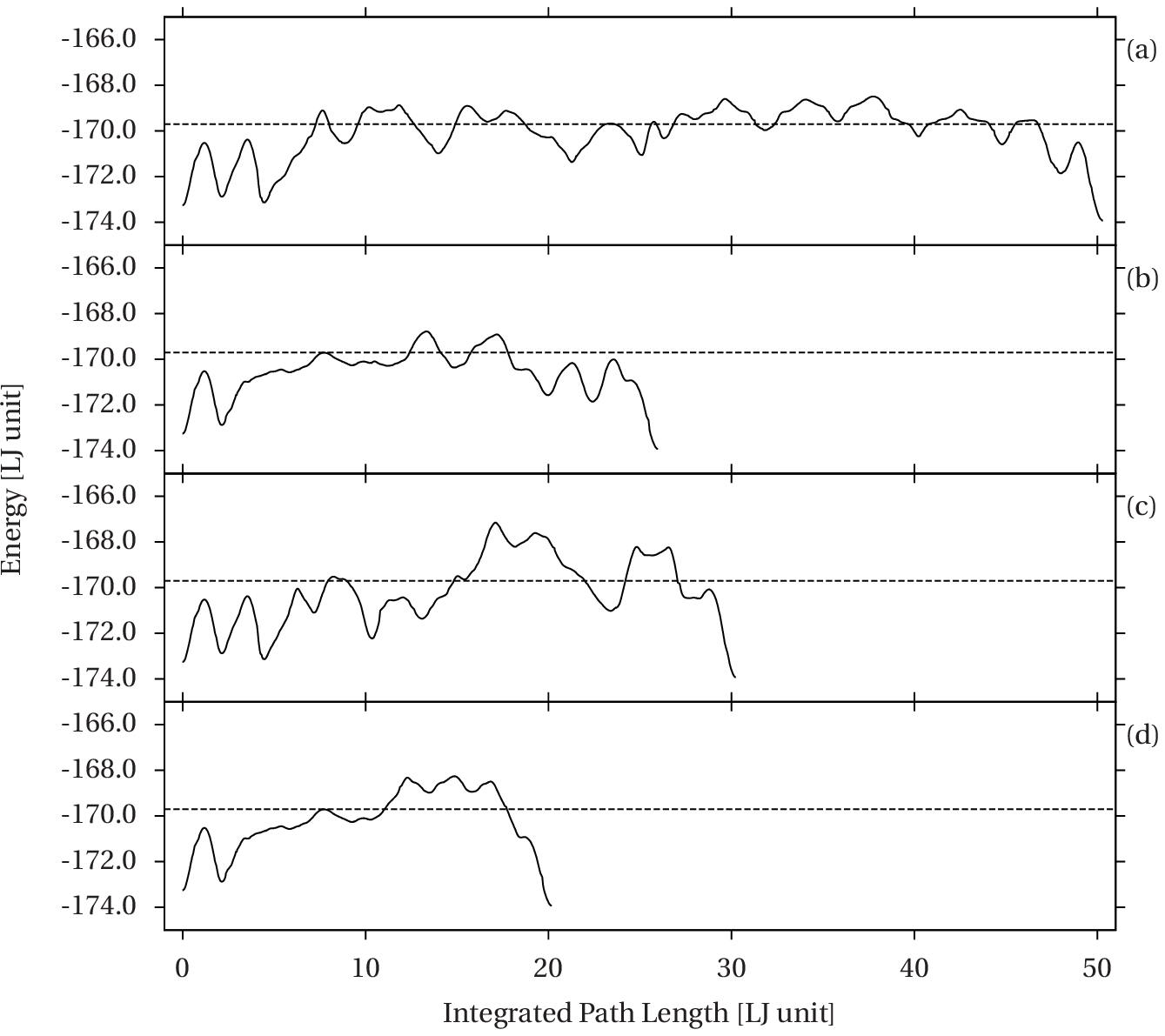}
  \caption{\label{fig:lj38fpinputpath}
Energy minimized pathways connecting the two lowest minima of
$\text{LJ}_{38}$. The pathway in
panel~(a) was obtained by extracting a sequence of minima from the
complete trajectory-based connectivity database. Panels~(b), (c)
and (d) show pathways that were obtained by successively removing
the highest energy transition along the lowest-barrier pathway from the
trajectory-based connectivity database.  Using the sequences of the
extracted minima as input for the MHGPS\cite{Schaefer2014} method, complete
pathways were reconstructed. The SQNS\cite{Schaefer2015} and
SQNM\cite{Schaefer2015} methods were used for converging to transition
states and relaxing to the connected minima.
}
\end{figure*}
Besides for generating disconnectivity graphs and qualitatively judging the
kinetics and thermodynamics of PESs, trajectory-based connectivity databases
can also be used to extract well aligned sequences of minima.  These well
aligned sequences of minima can be hoped to lie on a low-energy pathway between
two given states. Such minima sequences are of great importance, because they
provide promising starting points for generating initial pathways that are
needed for methods like TPS or its discrete variant,
DPS.\cite{Dellago1998-1,Dellago1998-2,Bolhuis2002,Dellago2003,Gruenwald2008,Gruenwald2009,Lechner2011,Wales2002,Wales2004-2}
For non-trivial reactions involving large structural changes such a generation
of initial pathways is in itself a very difficult task and no generally
applicable solution seems to exist, so far.\cite{Dellago2007} Isolating a
suitable sequence of minima from a trajectory-based connectivity database can
be done by applying a modified Dijkstra's algorithm which in a first round
searches for a path that minimizes the maximum barrier at any of its
transitions and in a second round minimizes with respect to the number of
intermediate transitions.\cite{Schaefer2014} Of course, the thus isolated
pathways are not necessarily complete in the sense that it might not be
possible to connect the two minima of a hopping pair by only one single
intermediate transition state.  However, the isolated sequence of minima
represents minima that were visited in consecutive order by an MH walker.
Therefore, they are suitable for getting connected by the connectivity finder
module of the MHGPS code (instead of the usual sequence of accepted MH
configurations).

For the $\text{Si}_{20}$ system a sequence of minima between the putative
global minimum and the putative second lowest minimum was extracted from the
trajectory-based connectivity database. For this sequence of minima, all
intermediate transition states and further emerging intermediate minima were
determined by means of the connectivity finder module of the MHGPS code as
implemented in the BigDFT suite. A pathway given by the trajectories of the
SQNM energy minimizer\cite{Schaefer2015} is shown in
Fig.~\ref{fig:si20fpinputpath}a. This pathway consists of 27 intermediate
transition states. Fig.~\ref{fig:si20fpinputpath}b shows a lowest barrier
pathway that was extracted from the stationary point database which was sampled
by means of unbiased MHGPS runs and already used for the standard
disconnectivity graphs in Fig.~\ref{fig:fptrees}a.3. The pathway in
Fig.~\ref{fig:si20fpinputpath}b consists of 20 intermediate transition states.
Remarkably, both paths exhibit the same highest energy transition state which
is highlighted by the red arrows in Fig.~\ref{fig:si20fpinputpath}. Still, the
path extracted from the stationary point database
(Fig.~\ref{fig:si20fpinputpath}b) is shorter than the path in
Fig.~\ref{fig:si20fpinputpath}a, both in terms of the integrated path length
and in terms of the number of intermediate transition states.

Of course, there is no guarantee that extracting a sequence of minima from a
trajectory-based connectivity database and connecting these minima by searching
intermediate transition states will result in a pathway that has the same
highest barrier as the pathway with the lowest highest barrier that is
contained in a thoroughly sampled stationary point database. However, computer
experiments performed for the $\text{LJ}_{38}$ cluster indicate that physically
reasonable pathways can be extracted from trajectory-based connectivity
databases.  Using the modified Dijkstra's algorithm, a sequence of minima was
extracted from the complete trajectory-based connectivity database for
$\text{LJ}_{38}$. By successively removing the highest energy transition  along
the lowest barrier pathway from the trajectory-based connectivity database,
this process was repeated four more times. In this way, five different
sequences of minima were obtained. Again, for each sequence, missing
intermediate minima and transition states were added by means of the
connectivity finder module of the MHGPS code. This procedure resulted in four
pathways with non-identical highest barriers, which are shown in
Fig.~\ref{fig:lj38fpinputpath}. The dashed line at an energy of $-169.708$ LJ
units indicates the highest barrier along the lowest-known barrier
pathway.\cite{Doye1999,Doye1999-2} Considering the fact that, for instance, in
the case of argon 1~LJ energy unit corresponds to roughly
10~meV,\cite{Rahman1964,Rowley1975,White1999} one sees that the highest
barriers along the pathways in Fig.~\ref{fig:lj38fpinputpath} are not much
higher than this lowest-known barrier.

\section{Conclusion}
Based on Lennard-Jones, Silicon, Sodium-Chloride and Gold clusters, it was
found that uphill barrier energies of transition states between directly
connected minima tend to increase with increasing structural differences of the
two minima. Based on this insight it also turned out that post-processing MH
data at a negligible computational cost allows to obtain qualitative
topological information on PESs that is stored in so called trajectory-based
connectivity database. These trajectory-based connectivity database can be used
for generating fingerprint disconnectivity graphs that allow to obtain a
qualitative idea on thermodynamic and kinetic properties of a system of
interest.  Besides allowing to asses system properties without the need of a
computational expensive explicit sampling of transition states and the
assessment of the PES's connectivity based on minimum energy or energy
minimized pathways, this method also serves as a valuable tool that can help to
decide if a certain multi-atomic system may exhibit desired properties before
investing significant resources for assessing theses properties more
rigorously.  Furthermore it was demonstrated that it is possible to extract
from a trajectory-based connectivity database well aligned sequences of minima
that can be used to generate initial pathways that are needed for methods like
TPS or DPS.

\section*{Supplementary Material}
See the supplementary material for the Gaussian kernel density estimates of
the uphill barrier energies versus the \textit{s}-orbital fingerprint
distances and the fingerprint disconnectivity graphs based on
\textit{s}-orbitals, only.
\begin{acknowledgments}
This research was supported by the NCCR MARVEL, funded by the Swiss National Science Foundation. Computer
time was provided by the Swiss National Supercomputing
Centre (CSCS) under project ID s499.

\end{acknowledgments}


\bibliography{fingerprinttree.bib}

\begin{thebibliography}{66}%
\makeatletter
\providecommand \@ifxundefined [1]{%
 \@ifx{#1\undefined}
}%
\providecommand \@ifnum [1]{%
 \ifnum #1\expandafter \@firstoftwo
 \else \expandafter \@secondoftwo
 \fi
}%
\providecommand \@ifx [1]{%
 \ifx #1\expandafter \@firstoftwo
 \else \expandafter \@secondoftwo
 \fi
}%
\providecommand \natexlab [1]{#1}%
\providecommand \enquote  [1]{``#1''}%
\providecommand \bibnamefont  [1]{#1}%
\providecommand \bibfnamefont [1]{#1}%
\providecommand \citenamefont [1]{#1}%
\providecommand \href@noop [0]{\@secondoftwo}%
\providecommand \href [0]{\begingroup \@sanitize@url \@href}%
\providecommand \@href[1]{\@@startlink{#1}\@@href}%
\providecommand \@@href[1]{\endgroup#1\@@endlink}%
\providecommand \@sanitize@url [0]{\catcode `\\12\catcode `\$12\catcode
  `\&12\catcode `\#12\catcode `\^12\catcode `\_12\catcode `\%12\relax}%
\providecommand \@@startlink[1]{}%
\providecommand \@@endlink[0]{}%
\providecommand \url  [0]{\begingroup\@sanitize@url \@url }%
\providecommand \@url [1]{\endgroup\@href {#1}{\urlprefix }}%
\providecommand \urlprefix  [0]{URL }%
\providecommand \Eprint [0]{\href }%
\providecommand \doibase [0]{http://dx.doi.org/}%
\providecommand \selectlanguage [0]{\@gobble}%
\providecommand \bibinfo  [0]{\@secondoftwo}%
\providecommand \bibfield  [0]{\@secondoftwo}%
\providecommand \translation [1]{[#1]}%
\providecommand \BibitemOpen [0]{}%
\providecommand \bibitemStop [0]{}%
\providecommand \bibitemNoStop [0]{.\EOS\space}%
\providecommand \EOS [0]{\spacefactor3000\relax}%
\providecommand \BibitemShut  [1]{\csname bibitem#1\endcsname}%
\let\auto@bib@innerbib\@empty
\bibitem [{\citenamefont {{L}evinthal}(1969)}]{Levinthal1969}%
  \BibitemOpen
  \bibfield  {author} {\bibinfo {author} {\bibfnamefont {C.}~\bibnamefont
  {{L}evinthal}},\ }in\ \href@noop {} {\emph {\bibinfo {booktitle}
  {{M}\"{o}ssbauer {S}pectroscopy in {B}iological {S}ystems: {P}roceedings of a
  {M}eeting held at {A}llerton {H}ouse, {M}onticello, {I}llinois}}},\ \bibinfo
  {editor} {edited by\ \bibinfo {editor} {\bibfnamefont {P.}~\bibnamefont
  {Debrunner}}, \bibinfo {editor} {\bibfnamefont {J.~C.~M.}\ \bibnamefont
  {Tsibris}}, \ and\ \bibinfo {editor} {\bibfnamefont {E.}~\bibnamefont
  {M\"{u}nck}}}\ (\bibinfo  {publisher} {University of Illinois Press},\
  \bibinfo {address} {Urbana},\ \bibinfo {year} {1969})\ pp.\ \bibinfo {pages}
  {22--24}\BibitemShut {NoStop}%
\bibitem [{\citenamefont {{D}ill}\ and\ \citenamefont
  {{C}han}(1997)}]{Dill1997}%
  \BibitemOpen
  \bibfield  {author} {\bibinfo {author} {\bibfnamefont {K.~A.}\ \bibnamefont
  {{D}ill}}\ and\ \bibinfo {author} {\bibfnamefont {H.~S.}\ \bibnamefont
  {{C}han}},\ }\href@noop {} {\bibfield  {journal} {\bibinfo  {journal} {Nature
  Structural Biology}\ }\textbf {\bibinfo {volume} {4}},\ \bibinfo {pages} {10}
  (\bibinfo {year} {1997})}\BibitemShut {NoStop}%
\bibitem [{\citenamefont {{D}avid {W}ales}(2003)}]{Wales2003}%
  \BibitemOpen
  \bibfield  {author} {\bibinfo {author} {\bibnamefont {{D}avid {W}ales}},\
  }\href@noop {} {\emph {\bibinfo {title} {{E}nergy {L}andscapes:
  {A}pplications to {C}lusters, {B}iomolecules and {G}lasses}}}\ (\bibinfo
  {publisher} {Cambridge University Press},\ \bibinfo {address} {Cambridge},\
  \bibinfo {year} {2003})\BibitemShut {NoStop}%
\bibitem [{\citenamefont {{D}ellago}\ \emph {et~al.}(1998)\citenamefont
  {{D}ellago}, \citenamefont {{B}olhuis}, \citenamefont {{C}sajka},\ and\
  \citenamefont {{C}handler}}]{Dellago1998-1}%
  \BibitemOpen
  \bibfield  {author} {\bibinfo {author} {\bibfnamefont {C.}~\bibnamefont
  {{D}ellago}}, \bibinfo {author} {\bibfnamefont {P.~G.}\ \bibnamefont
  {{B}olhuis}}, \bibinfo {author} {\bibfnamefont {F.~S.}\ \bibnamefont
  {{C}sajka}}, \ and\ \bibinfo {author} {\bibfnamefont {D.}~\bibnamefont
  {{C}handler}},\ }\href@noop {} {\bibfield  {journal} {\bibinfo  {journal}
  {The Journal of Chemical Physics}\ }\textbf {\bibinfo {volume} {108}},\
  \bibinfo {pages} {1964} (\bibinfo {year} {1998})}\BibitemShut {NoStop}%
\bibitem [{\citenamefont {{D}ellago}, \citenamefont {{B}olhuis},\ and\
  \citenamefont {{C}handler}(1998)}]{Dellago1998-2}%
  \BibitemOpen
  \bibfield  {author} {\bibinfo {author} {\bibfnamefont {C.}~\bibnamefont
  {{D}ellago}}, \bibinfo {author} {\bibfnamefont {P.~G.}\ \bibnamefont
  {{B}olhuis}}, \ and\ \bibinfo {author} {\bibfnamefont {D.}~\bibnamefont
  {{C}handler}},\ }\href@noop {} {\bibfield  {journal} {\bibinfo  {journal}
  {The Journal of Chemical Physics}\ }\textbf {\bibinfo {volume} {108}},\
  \bibinfo {pages} {9236} (\bibinfo {year} {1998})}\BibitemShut {NoStop}%
\bibitem [{\citenamefont {{B}olhuis}\ \emph {et~al.}(2002)\citenamefont
  {{B}olhuis}, \citenamefont {{C}handler}, \citenamefont {{D}ellago},\ and\
  \citenamefont {{G}eissler}}]{Bolhuis2002}%
  \BibitemOpen
  \bibfield  {author} {\bibinfo {author} {\bibfnamefont {P.~G.}\ \bibnamefont
  {{B}olhuis}}, \bibinfo {author} {\bibfnamefont {D.}~\bibnamefont
  {{C}handler}}, \bibinfo {author} {\bibfnamefont {C.}~\bibnamefont
  {{D}ellago}}, \ and\ \bibinfo {author} {\bibfnamefont {P.~L.}\ \bibnamefont
  {{G}eissler}},\ }\href@noop {} {\bibfield  {journal} {\bibinfo  {journal}
  {Annual Review of Physical Chemistry}\ }\textbf {\bibinfo {volume} {53}},\
  \bibinfo {pages} {291} (\bibinfo {year} {2002})}\BibitemShut {NoStop}%
\bibitem [{\citenamefont {{D}ellago}, \citenamefont {{B}olhuis},\ and\
  \citenamefont {{G}eissler}(2003)}]{Dellago2003}%
  \BibitemOpen
  \bibfield  {author} {\bibinfo {author} {\bibfnamefont {C.}~\bibnamefont
  {{D}ellago}}, \bibinfo {author} {\bibfnamefont {P.~G.}\ \bibnamefont
  {{B}olhuis}}, \ and\ \bibinfo {author} {\bibfnamefont {P.~L.}\ \bibnamefont
  {{G}eissler}},\ }\enquote {\bibinfo {title} {{T}ransition {P}ath
  {S}ampling},}\ in\ \href@noop {} {\emph {\bibinfo {booktitle} {Advances in
  Chemical Physics}}}\ (\bibinfo  {publisher} {John Wiley \& Sons, Inc.},\
  \bibinfo {year} {2003})\ pp.\ \bibinfo {pages} {1--78}\BibitemShut {NoStop}%
\bibitem [{\citenamefont {{G}r\"{u}nwald}, \citenamefont {{D}ellago},\ and\
  \citenamefont {{G}eissler}(2008)}]{Gruenwald2008}%
  \BibitemOpen
  \bibfield  {author} {\bibinfo {author} {\bibfnamefont {M.}~\bibnamefont
  {{G}r\"{u}nwald}}, \bibinfo {author} {\bibfnamefont {C.}~\bibnamefont
  {{D}ellago}}, \ and\ \bibinfo {author} {\bibfnamefont {P.~L.}\ \bibnamefont
  {{G}eissler}},\ }\href@noop {} {\bibfield  {journal} {\bibinfo  {journal}
  {The Journal of Chemical Physics}\ }\textbf {\bibinfo {volume} {129}},\
  \bibinfo {pages} {194101} (\bibinfo {year} {2008})}\BibitemShut {NoStop}%
\bibitem [{\citenamefont {{G}r\"{u}nwald}\ and\ \citenamefont
  {{D}ellago}(2009)}]{Gruenwald2009}%
  \BibitemOpen
  \bibfield  {author} {\bibinfo {author} {\bibfnamefont {M.}~\bibnamefont
  {{G}r\"{u}nwald}}\ and\ \bibinfo {author} {\bibfnamefont {C.}~\bibnamefont
  {{D}ellago}},\ }\href@noop {} {\bibfield  {journal} {\bibinfo  {journal}
  {Nano Letters}\ }\textbf {\bibinfo {volume} {9}},\ \bibinfo {pages} {2099}
  (\bibinfo {year} {2009})}\BibitemShut {NoStop}%
\bibitem [{\citenamefont {{L}echner}, \citenamefont {{D}ellago},\ and\
  \citenamefont {{B}olhuis}(2011)}]{Lechner2011}%
  \BibitemOpen
  \bibfield  {author} {\bibinfo {author} {\bibfnamefont {W.}~\bibnamefont
  {{L}echner}}, \bibinfo {author} {\bibfnamefont {C.}~\bibnamefont
  {{D}ellago}}, \ and\ \bibinfo {author} {\bibfnamefont {P.~G.}\ \bibnamefont
  {{B}olhuis}},\ }\href@noop {} {\bibfield  {journal} {\bibinfo  {journal}
  {Physical Review Letters}\ }\textbf {\bibinfo {volume} {106}},\ \bibinfo
  {pages} {085701} (\bibinfo {year} {2011})}\BibitemShut {NoStop}%
\bibitem [{\citenamefont {{D}avid {J}.~{W}ales}(2002)}]{Wales2002}%
  \BibitemOpen
  \bibfield  {author} {\bibinfo {author} {\bibnamefont {{D}avid
  {J}.~{W}ales}},\ }\href@noop {} {\bibfield  {journal} {\bibinfo  {journal}
  {Molecular Physics}\ }\textbf {\bibinfo {volume} {100}},\ \bibinfo {pages}
  {3285} (\bibinfo {year} {2002})}\BibitemShut {NoStop}%
\bibitem [{\citenamefont {{W}ales}(2004)}]{Wales2004-2}%
  \BibitemOpen
  \bibfield  {author} {\bibinfo {author} {\bibfnamefont {D.~J.}\ \bibnamefont
  {{W}ales}},\ }\href@noop {} {\bibfield  {journal} {\bibinfo  {journal}
  {Molecular Physics}\ }\textbf {\bibinfo {volume} {102}},\ \bibinfo {pages}
  {891} (\bibinfo {year} {2004})}\BibitemShut {NoStop}%
\bibitem [{\citenamefont {{Z}hang}\ and\ \citenamefont
  {{L}iu}(2015)}]{Zhang2015}%
  \BibitemOpen
  \bibfield  {author} {\bibinfo {author} {\bibfnamefont {X.-J.}\ \bibnamefont
  {{Z}hang}}\ and\ \bibinfo {author} {\bibfnamefont {Z.-P.}\ \bibnamefont
  {{L}iu}},\ }\href@noop {} {\bibfield  {journal} {\bibinfo  {journal}
  {Physical Chemistry Chemical Physics}\ }\textbf {\bibinfo {volume} {17}},\
  \bibinfo {pages} {2757} (\bibinfo {year} {2015})}\BibitemShut {NoStop}%
\bibitem [{\citenamefont {{M}ousseau}\ and\ \citenamefont
  {{B}arkema}(1998)}]{Mousseau1998}%
  \BibitemOpen
  \bibfield  {author} {\bibinfo {author} {\bibfnamefont {N.}~\bibnamefont
  {{M}ousseau}}\ and\ \bibinfo {author} {\bibfnamefont {G.~T.}\ \bibnamefont
  {{B}arkema}},\ }\href@noop {} {\bibfield  {journal} {\bibinfo  {journal}
  {Physical Review E}\ }\textbf {\bibinfo {volume} {57}},\ \bibinfo {pages}
  {2419} (\bibinfo {year} {1998})}\BibitemShut {NoStop}%
\bibitem [{\citenamefont {{W}ei}, \citenamefont {{M}ousseau},\ and\
  \citenamefont {{D}erreumaux}(2002)}]{Wei2002}%
  \BibitemOpen
  \bibfield  {author} {\bibinfo {author} {\bibfnamefont {G.}~\bibnamefont
  {{W}ei}}, \bibinfo {author} {\bibfnamefont {N.}~\bibnamefont {{M}ousseau}}, \
  and\ \bibinfo {author} {\bibfnamefont {P.}~\bibnamefont {{D}erreumaux}},\
  }\href@noop {} {\bibfield  {journal} {\bibinfo  {journal} {The Journal of
  Chemical Physics}\ }\textbf {\bibinfo {volume} {117}},\ \bibinfo {pages}
  {11379} (\bibinfo {year} {2002})}\BibitemShut {NoStop}%
\bibitem [{\citenamefont {{M}achado {C}harry}\ \emph
  {et~al.}(2011)\citenamefont {{M}achado {C}harry}, \citenamefont
  {{B}\'{e}land}, \citenamefont {{C}aliste}, \citenamefont {{G}enovese},
  \citenamefont {{D}eutsch}, \citenamefont {{M}ousseau},\ and\ \citenamefont
  {{P}ochet}}]{Machado2011}%
  \BibitemOpen
  \bibfield  {author} {\bibinfo {author} {\bibfnamefont {E.}~\bibnamefont
  {{M}achado {C}harry}}, \bibinfo {author} {\bibfnamefont {L.~K.}\ \bibnamefont
  {{B}\'{e}land}}, \bibinfo {author} {\bibfnamefont {D.}~\bibnamefont
  {{C}aliste}}, \bibinfo {author} {\bibfnamefont {L.}~\bibnamefont
  {{G}enovese}}, \bibinfo {author} {\bibfnamefont {T.}~\bibnamefont
  {{D}eutsch}}, \bibinfo {author} {\bibfnamefont {N.}~\bibnamefont
  {{M}ousseau}}, \ and\ \bibinfo {author} {\bibfnamefont {P.}~\bibnamefont
  {{P}ochet}},\ }\href@noop {} {\bibfield  {journal} {\bibinfo  {journal} {The
  Journal of Chemical Physics}\ }\textbf {\bibinfo {volume} {135}},\ \bibinfo
  {pages} {34102} (\bibinfo {year} {2011})}\BibitemShut {NoStop}%
\bibitem [{\citenamefont {{M}ousseau}\ \emph {et~al.}(2012)\citenamefont
  {{M}ousseau}, \citenamefont {{B}\'{e}land}, \citenamefont {{B}rommer},
  \citenamefont {{J}oly}, \citenamefont {{E}l {M}ellouhi}, \citenamefont
  {{M}achado {C}harry}, \citenamefont {{M}arinica},\ and\ \citenamefont
  {{P}ochet}}]{Mousseau2012}%
  \BibitemOpen
  \bibfield  {author} {\bibinfo {author} {\bibfnamefont {N.}~\bibnamefont
  {{M}ousseau}}, \bibinfo {author} {\bibfnamefont {L.~K.}\ \bibnamefont
  {{B}\'{e}land}}, \bibinfo {author} {\bibfnamefont {P.}~\bibnamefont
  {{B}rommer}}, \bibinfo {author} {\bibfnamefont {J.-F.}\ \bibnamefont
  {{J}oly}}, \bibinfo {author} {\bibfnamefont {F.}~\bibnamefont {{E}l
  {M}ellouhi}}, \bibinfo {author} {\bibfnamefont {E.}~\bibnamefont {{M}achado
  {C}harry}}, \bibinfo {author} {\bibfnamefont {M.-C.}\ \bibnamefont
  {{M}arinica}}, \ and\ \bibinfo {author} {\bibfnamefont {P.}~\bibnamefont
  {{P}ochet}},\ }\href@noop {} {\bibfield  {journal} {\bibinfo  {journal}
  {Journal of Atomic and Molecular Physics}\ }\textbf {\bibinfo {volume}
  {2012}},\ \bibinfo {pages} {925278} (\bibinfo {year} {2012})}\BibitemShut
  {NoStop}%
\bibitem [{\citenamefont {{S}{\o}rensen}\ and\ \citenamefont
  {{V}oter}(2000)}]{Sorensen2000}%
  \BibitemOpen
  \bibfield  {author} {\bibinfo {author} {\bibfnamefont {M.~R.}\ \bibnamefont
  {{S}{\o}rensen}}\ and\ \bibinfo {author} {\bibfnamefont {A.~F.}\ \bibnamefont
  {{V}oter}},\ }\href@noop {} {\bibfield  {journal} {\bibinfo  {journal} {The
  Journal of Chemical Physics}\ }\textbf {\bibinfo {volume} {112}},\ \bibinfo
  {pages} {9599} (\bibinfo {year} {2000})}\BibitemShut {NoStop}%
\bibitem [{\citenamefont {{D}anny {P}erez}\ \emph {et~al.}(2009)\citenamefont
  {{D}anny {P}erez}, \citenamefont {{B}las {P}.~{U}beruaga}, \citenamefont
  {{Y}unsic {S}him}, \citenamefont {{J}acques {G}.~{A}mar},\ and\ \citenamefont
  {{A}rthur {F}.~{V}oter}}]{Perez2009}%
  \BibitemOpen
  \bibfield  {author} {\bibinfo {author} {\bibnamefont {{D}anny {P}erez}},
  \bibinfo {author} {\bibnamefont {{B}las {P}.~{U}beruaga}}, \bibinfo {author}
  {\bibnamefont {{Y}unsic {S}him}}, \bibinfo {author} {\bibnamefont {{J}acques
  {G}.~{A}mar}}, \ and\ \bibinfo {author} {\bibnamefont {{A}rthur
  {F}.~{V}oter}}\ }(\bibinfo  {publisher} {{E}lsevier},\ \bibinfo {year}
  {2009})\ pp.\ \bibinfo {pages} {79 -- 98}\BibitemShut {NoStop}%
\bibitem [{\citenamefont {{G}oedecker}(2004)}]{Goedecker2004}%
  \BibitemOpen
  \bibfield  {author} {\bibinfo {author} {\bibfnamefont {S.}~\bibnamefont
  {{G}oedecker}},\ }\href@noop {} {\bibfield  {journal} {\bibinfo  {journal}
  {The Journal of Chemical Physics}\ }\textbf {\bibinfo {volume} {120}},\
  \bibinfo {pages} {9911} (\bibinfo {year} {2004})}\BibitemShut {NoStop}%
\bibitem [{\citenamefont {{S}chaefer}\ \emph
  {et~al.}(2014{\natexlab{a}})\citenamefont {{S}chaefer}, \citenamefont
  {{M}ohr}, \citenamefont {{A}msler},\ and\ \citenamefont
  {{G}oedecker}}]{Schaefer2014}%
  \BibitemOpen
  \bibfield  {author} {\bibinfo {author} {\bibfnamefont {B.}~\bibnamefont
  {{S}chaefer}}, \bibinfo {author} {\bibfnamefont {S.}~\bibnamefont {{M}ohr}},
  \bibinfo {author} {\bibfnamefont {M.}~\bibnamefont {{A}msler}}, \ and\
  \bibinfo {author} {\bibfnamefont {S.}~\bibnamefont {{G}oedecker}},\
  }\href@noop {} {\bibfield  {journal} {\bibinfo  {journal} {The Journal of
  Chemical Physics}\ }\textbf {\bibinfo {volume} {140}},\ \bibinfo {pages}
  {214102} (\bibinfo {year} {2014}{\natexlab{a}})}\BibitemShut {NoStop}%
\bibitem [{\citenamefont {{S}chaefer}\ \emph {et~al.}(2015)\citenamefont
  {{S}chaefer}, \citenamefont {{G}hasemi}, \citenamefont {{R}oy},\ and\
  \citenamefont {{G}oedecker}}]{Schaefer2015}%
  \BibitemOpen
  \bibfield  {author} {\bibinfo {author} {\bibfnamefont {B.}~\bibnamefont
  {{S}chaefer}}, \bibinfo {author} {\bibfnamefont {S.~A.}\ \bibnamefont
  {{G}hasemi}}, \bibinfo {author} {\bibfnamefont {S.}~\bibnamefont {{R}oy}}, \
  and\ \bibinfo {author} {\bibfnamefont {S.}~\bibnamefont {{G}oedecker}},\
  }\href@noop {} {\bibfield  {journal} {\bibinfo  {journal} {The Journal of
  Chemical Physics}\ }\textbf {\bibinfo {volume} {142}},\ \bibinfo {pages}
  {034112} (\bibinfo {year} {2015})}\BibitemShut {NoStop}%
\bibitem [{\citenamefont {{A}msler}\ and\ \citenamefont
  {{G}oedecker}(2010)}]{Amsler2010}%
  \BibitemOpen
  \bibfield  {author} {\bibinfo {author} {\bibfnamefont {M.}~\bibnamefont
  {{A}msler}}\ and\ \bibinfo {author} {\bibfnamefont {S.}~\bibnamefont
  {{G}oedecker}},\ }\href@noop {} {\bibfield  {journal} {\bibinfo  {journal}
  {The Journal of Chemical Physics}\ }\textbf {\bibinfo {volume} {133}},\
  \bibinfo {pages} {224104} (\bibinfo {year} {2010})}\BibitemShut {NoStop}%
\bibitem [{\citenamefont {{O}ganov}\ and\ \citenamefont
  {{G}lass}(2006)}]{Oganov2006}%
  \BibitemOpen
  \bibfield  {author} {\bibinfo {author} {\bibfnamefont {A.~R.}\ \bibnamefont
  {{O}ganov}}\ and\ \bibinfo {author} {\bibfnamefont {C.~W.}\ \bibnamefont
  {{G}lass}},\ }\href@noop {} {\bibfield  {journal} {\bibinfo  {journal} {The
  Journal of Chemical Physics}\ }\textbf {\bibinfo {volume} {124}},\ \bibinfo
  {pages} {244704} (\bibinfo {year} {2006})}\BibitemShut {NoStop}%
\bibitem [{\citenamefont {{C}olin {W}.~{G}lass}, \citenamefont {{A}rtem
  {R}.~{O}ganov},\ and\ \citenamefont {{N}ikolaus {H}ansen}(2006)}]{Glass2006}%
  \BibitemOpen
  \bibfield  {author} {\bibinfo {author} {\bibnamefont {{C}olin {W}.~{G}lass}},
  \bibinfo {author} {\bibnamefont {{A}rtem {R}.~{O}ganov}}, \ and\ \bibinfo
  {author} {\bibnamefont {{N}ikolaus {H}ansen}},\ }\href@noop {} {\bibfield
  {journal} {\bibinfo  {journal} {Computer Physics Communications}\ }\textbf
  {\bibinfo {volume} {175}},\ \bibinfo {pages} {713 } (\bibinfo {year}
  {2006})}\BibitemShut {NoStop}%
\bibitem [{\citenamefont {{W}ales}\ and\ \citenamefont
  {{D}oye}(1997)}]{Wales1997}%
  \BibitemOpen
  \bibfield  {author} {\bibinfo {author} {\bibfnamefont {D.~J.}\ \bibnamefont
  {{W}ales}}\ and\ \bibinfo {author} {\bibfnamefont {J.~P.~K.}\ \bibnamefont
  {{D}oye}},\ }\href@noop {} {\bibfield  {journal} {\bibinfo  {journal}
  {Journal of Physical Chemistry A}\ }\textbf {\bibinfo {volume} {101}},\
  \bibinfo {pages} {5111} (\bibinfo {year} {1997})}\BibitemShut {NoStop}%
\bibitem [{\citenamefont {{D}oye}\ and\ \citenamefont
  {{W}ales}(1998)}]{Doye1998}%
  \BibitemOpen
  \bibfield  {author} {\bibinfo {author} {\bibfnamefont {J.~P.~K.}\
  \bibnamefont {{D}oye}}\ and\ \bibinfo {author} {\bibfnamefont {D.~J.}\
  \bibnamefont {{W}ales}},\ }\href@noop {} {\bibfield  {journal} {\bibinfo
  {journal} {Phys. Rev. Lett.}\ }\textbf {\bibinfo {volume} {80}},\ \bibinfo
  {pages} {1357} (\bibinfo {year} {1998})}\BibitemShut {NoStop}%
\bibitem [{\citenamefont {{D}oye}, \citenamefont {{W}ales},\ and\ \citenamefont
  {{M}iller}(1998)}]{Doye1998b}%
  \BibitemOpen
  \bibfield  {author} {\bibinfo {author} {\bibfnamefont {J.~P.~K.}\
  \bibnamefont {{D}oye}}, \bibinfo {author} {\bibfnamefont {D.~J.}\
  \bibnamefont {{W}ales}}, \ and\ \bibinfo {author} {\bibfnamefont {M.~A.}\
  \bibnamefont {{M}iller}},\ }\href@noop {} {\bibfield  {journal} {\bibinfo
  {journal} {The Journal of Chemical Physics}\ }\textbf {\bibinfo {volume}
  {109}},\ \bibinfo {pages} {8143} (\bibinfo {year} {1998})}\BibitemShut
  {NoStop}%
\bibitem [{\citenamefont {{P}ickard}\ and\ \citenamefont
  {{N}eeds}(2006)}]{Pickard2006}%
  \BibitemOpen
  \bibfield  {author} {\bibinfo {author} {\bibfnamefont {C.~J.}\ \bibnamefont
  {{P}ickard}}\ and\ \bibinfo {author} {\bibfnamefont {R.~J.}\ \bibnamefont
  {{N}eeds}},\ }\href@noop {} {\bibfield  {journal} {\bibinfo  {journal}
  {Physical Review Letters}\ }\textbf {\bibinfo {volume} {97}},\ \bibinfo
  {pages} {045504} (\bibinfo {year} {2006})}\BibitemShut {NoStop}%
\bibitem [{\citenamefont {{P}ickard}\ and\ \citenamefont
  {{N}eeds}(2007{\natexlab{a}})}]{Pickard2007}%
  \BibitemOpen
  \bibfield  {author} {\bibinfo {author} {\bibfnamefont {C.~J.}\ \bibnamefont
  {{P}ickard}}\ and\ \bibinfo {author} {\bibfnamefont {R.~J.}\ \bibnamefont
  {{N}eeds}},\ }\href@noop {} {\bibfield  {journal} {\bibinfo  {journal}
  {Nature Physics}\ }\textbf {\bibinfo {volume} {3}},\ \bibinfo {pages} {473}
  (\bibinfo {year} {2007}{\natexlab{a}})}\BibitemShut {NoStop}%
\bibitem [{\citenamefont {{P}ickard}\ and\ \citenamefont
  {{N}eeds}(2007{\natexlab{b}})}]{Pickard2007b}%
  \BibitemOpen
  \bibfield  {author} {\bibinfo {author} {\bibfnamefont {C.~J.}\ \bibnamefont
  {{P}ickard}}\ and\ \bibinfo {author} {\bibfnamefont {R.~J.}\ \bibnamefont
  {{N}eeds}},\ }\href@noop {} {\bibfield  {journal} {\bibinfo  {journal}
  {Physical Review B}\ }\textbf {\bibinfo {volume} {76}},\ \bibinfo {pages}
  {144114} (\bibinfo {year} {2007}{\natexlab{b}})}\BibitemShut {NoStop}%
\bibitem [{\citenamefont {{P}ickard}\ and\ \citenamefont
  {{N}eeds}()}]{Pickard2008}%
  \BibitemOpen
  \bibfield  {author} {\bibinfo {author} {\bibfnamefont {C.~J.}\ \bibnamefont
  {{P}ickard}}\ and\ \bibinfo {author} {\bibfnamefont {R.~J.}\ \bibnamefont
  {{N}eeds}},\ }\href@noop {} {\bibinfo  {journal} {Nature Materials}\ ,\
  \bibinfo {pages} {775}}\BibitemShut {NoStop}%
\bibitem [{\citenamefont {{D}e}\ \emph {et~al.}(2014)\citenamefont {{D}e},
  \citenamefont {{S}chaefer}, \citenamefont {{S}adeghi}, \citenamefont
  {{S}icher}, \citenamefont {{K}anhere},\ and\ \citenamefont
  {{G}oedecker}}]{De2014}%
  \BibitemOpen
\bibfield  {journal} {  }\bibfield  {author} {\bibinfo {author} {\bibfnamefont
  {S.}~\bibnamefont {{D}e}}, \bibinfo {author} {\bibfnamefont {B.}~\bibnamefont
  {{S}chaefer}}, \bibinfo {author} {\bibfnamefont {A.}~\bibnamefont
  {{S}adeghi}}, \bibinfo {author} {\bibfnamefont {M.}~\bibnamefont {{S}icher}},
  \bibinfo {author} {\bibfnamefont {D.~G.}\ \bibnamefont {{K}anhere}}, \ and\
  \bibinfo {author} {\bibfnamefont {S.}~\bibnamefont {{G}oedecker}},\
  }\href@noop {} {\bibfield  {journal} {\bibinfo  {journal} {Physical Review
  Letters}\ }\textbf {\bibinfo {volume} {112}},\ \bibinfo {pages} {083401}
  (\bibinfo {year} {2014})}\BibitemShut {NoStop}%
\bibitem [{\citenamefont {{S}adeghi}\ \emph {et~al.}(2013)\citenamefont
  {{S}adeghi}, \citenamefont {{G}hasemi}, \citenamefont {{S}chaefer},
  \citenamefont {{M}ohr}, \citenamefont {{L}ill},\ and\ \citenamefont
  {{G}oedecker}}]{Sadeghi2013}%
  \BibitemOpen
  \bibfield  {author} {\bibinfo {author} {\bibfnamefont {A.}~\bibnamefont
  {{S}adeghi}}, \bibinfo {author} {\bibfnamefont {S.~A.}\ \bibnamefont
  {{G}hasemi}}, \bibinfo {author} {\bibfnamefont {B.}~\bibnamefont
  {{S}chaefer}}, \bibinfo {author} {\bibfnamefont {S.}~\bibnamefont {{M}ohr}},
  \bibinfo {author} {\bibfnamefont {M.~A.}\ \bibnamefont {{L}ill}}, \ and\
  \bibinfo {author} {\bibfnamefont {S.}~\bibnamefont {{G}oedecker}},\
  }\href@noop {} {\bibfield  {journal} {\bibinfo  {journal} {The Journal of
  Chemical Physics}\ }\textbf {\bibinfo {volume} {139}},\ \bibinfo {pages}
  {184118} (\bibinfo {year} {2013})}\BibitemShut {NoStop}%
\bibitem [{\citenamefont {{O}ren {M}.~{B}ecker}\ and\ \citenamefont {{M}artin
  {K}arplus}(1997)}]{Becker1997}%
  \BibitemOpen
  \bibfield  {author} {\bibinfo {author} {\bibnamefont {{O}ren {M}.~{B}ecker}}\
  and\ \bibinfo {author} {\bibnamefont {{M}artin {K}arplus}},\ }\href@noop {}
  {\bibfield  {journal} {\bibinfo  {journal} {The Journal of Chemical Physics}\
  }\textbf {\bibinfo {volume} {106}},\ \bibinfo {pages} {1495} (\bibinfo {year}
  {1997})}\BibitemShut {NoStop}%
\bibitem [{\citenamefont {{G}enovese}\ \emph {et~al.}(2008)\citenamefont
  {{G}enovese}, \citenamefont {{N}eelov}, \citenamefont {{G}oedecker},
  \citenamefont {{D}eutsch}, \citenamefont {{G}hasemi}, \citenamefont
  {{W}illand}, \citenamefont {{C}aliste}, \citenamefont {{Z}ilberberg},
  \citenamefont {{R}ayson}, \citenamefont {{B}ergman},\ and\ \citenamefont
  {et~al.}}]{Genovese2008}%
  \BibitemOpen
  \bibfield  {author} {\bibinfo {author} {\bibfnamefont {L.}~\bibnamefont
  {{G}enovese}}, \bibinfo {author} {\bibfnamefont {A.}~\bibnamefont
  {{N}eelov}}, \bibinfo {author} {\bibfnamefont {S.}~\bibnamefont
  {{G}oedecker}}, \bibinfo {author} {\bibfnamefont {T.}~\bibnamefont
  {{D}eutsch}}, \bibinfo {author} {\bibfnamefont {S.~A.}\ \bibnamefont
  {{G}hasemi}}, \bibinfo {author} {\bibfnamefont {A.}~\bibnamefont
  {{W}illand}}, \bibinfo {author} {\bibfnamefont {D.}~\bibnamefont
  {{C}aliste}}, \bibinfo {author} {\bibfnamefont {O.}~\bibnamefont
  {{Z}ilberberg}}, \bibinfo {author} {\bibfnamefont {M.}~\bibnamefont
  {{R}ayson}}, \bibinfo {author} {\bibfnamefont {A.}~\bibnamefont {{B}ergman}},
  \ and\ \bibinfo {author} {\bibnamefont {et~al.}},\ }\href@noop {} {\bibfield
  {journal} {\bibinfo  {journal} {The Journal of Chemical Physics}\ }\textbf
  {\bibinfo {volume} {129}},\ \bibinfo {pages} {014109} (\bibinfo {year}
  {2008})}\BibitemShut {NoStop}%
\bibitem [{\citenamefont {{M}ohr}\ \emph {et~al.}(2014)\citenamefont {{M}ohr},
  \citenamefont {{R}atcliff}, \citenamefont {{B}oulanger}, \citenamefont
  {{G}enovese}, \citenamefont {{C}aliste}, \citenamefont {{D}eutsch},\ and\
  \citenamefont {{G}oedecker}}]{Mohr2014}%
  \BibitemOpen
  \bibfield  {author} {\bibinfo {author} {\bibfnamefont {S.}~\bibnamefont
  {{M}ohr}}, \bibinfo {author} {\bibfnamefont {L.~E.}\ \bibnamefont
  {{R}atcliff}}, \bibinfo {author} {\bibfnamefont {P.}~\bibnamefont
  {{B}oulanger}}, \bibinfo {author} {\bibfnamefont {L.}~\bibnamefont
  {{G}enovese}}, \bibinfo {author} {\bibfnamefont {D.}~\bibnamefont
  {{C}aliste}}, \bibinfo {author} {\bibfnamefont {T.}~\bibnamefont
  {{D}eutsch}}, \ and\ \bibinfo {author} {\bibfnamefont {S.}~\bibnamefont
  {{G}oedecker}},\ }\href {http://dx.doi.org/10.1063/1.4871876} {\bibfield
  {journal} {\bibinfo  {journal} {The Journal of Chemical Physics}\ }\textbf
  {\bibinfo {volume} {140}},\ \bibinfo {pages} {204110} (\bibinfo {year}
  {2014})}\BibitemShut {NoStop}%
\bibitem [{\citenamefont {{W}illand}\ \emph {et~al.}(2013)\citenamefont
  {{W}illand}, \citenamefont {{K}vashnin}, \citenamefont {{G}enovese},
  \citenamefont {{V}\'{a}zquez Mayagoitia}, \citenamefont {{D}eb},
  \citenamefont {{S}adeghi}, \citenamefont {{D}eutsch},\ and\ \citenamefont
  {{G}oedecker}}]{Willand2013}%
  \BibitemOpen
  \bibfield  {author} {\bibinfo {author} {\bibfnamefont {A.}~\bibnamefont
  {{W}illand}}, \bibinfo {author} {\bibfnamefont {Y.~O.}\ \bibnamefont
  {{K}vashnin}}, \bibinfo {author} {\bibfnamefont {L.}~\bibnamefont
  {{G}enovese}}, \bibinfo {author} {\bibfnamefont {{\'A}.}~\bibnamefont
  {{V}\'{a}zquez Mayagoitia}}, \bibinfo {author} {\bibfnamefont {A.~K.}\
  \bibnamefont {{D}eb}}, \bibinfo {author} {\bibfnamefont {A.}~\bibnamefont
  {{S}adeghi}}, \bibinfo {author} {\bibfnamefont {T.}~\bibnamefont
  {{D}eutsch}}, \ and\ \bibinfo {author} {\bibfnamefont {S.}~\bibnamefont
  {{G}oedecker}},\ }\href@noop {} {\bibfield  {journal} {\bibinfo  {journal}
  {The Journal of Chemical Physics}\ }\textbf {\bibinfo {volume} {138}},\
  \bibinfo {pages} {104109} (\bibinfo {year} {2013})}\BibitemShut {NoStop}%
\bibitem [{\citenamefont {{B}ell}(1936)}]{Bell1936}%
  \BibitemOpen
  \bibfield  {author} {\bibinfo {author} {\bibfnamefont {R.~P.}\ \bibnamefont
  {{B}ell}},\ }\href@noop {} {\bibfield  {journal} {\bibinfo  {journal}
  {Proceedings of the Royal Society of London A}\ }\textbf {\bibinfo {volume}
  {154}},\ \bibinfo {pages} {414} (\bibinfo {year} {1936})}\BibitemShut
  {NoStop}%
\bibitem [{\citenamefont {{E}vans}\ and\ \citenamefont
  {{P}olanyi}(1936)}]{Evans1936}%
  \BibitemOpen
  \bibfield  {author} {\bibinfo {author} {\bibfnamefont {M.~G.}\ \bibnamefont
  {{E}vans}}\ and\ \bibinfo {author} {\bibfnamefont {M.}~\bibnamefont
  {{P}olanyi}},\ }\href {http://dx.doi.org/10.1039/TF9363201333} {\bibfield
  {journal} {\bibinfo  {journal} {Trans. Faraday Soc.}\ }\textbf {\bibinfo
  {volume} {32}},\ \bibinfo {pages} {1333} (\bibinfo {year}
  {1936})}\BibitemShut {NoStop}%
\bibitem [{\citenamefont {{H}ammond}(1955)}]{Hammond1955}%
  \BibitemOpen
  \bibfield  {author} {\bibinfo {author} {\bibfnamefont {G.~S.}\ \bibnamefont
  {{H}ammond}},\ }\href@noop {} {\bibfield  {journal} {\bibinfo  {journal}
  {Journal of the American Chemical Society}\ }\textbf {\bibinfo {volume}
  {77}},\ \bibinfo {pages} {334} (\bibinfo {year} {1955})}\BibitemShut
  {NoStop}%
\bibitem [{\citenamefont {{M}arcus}(1968)}]{Marcus1968}%
  \BibitemOpen
  \bibfield  {author} {\bibinfo {author} {\bibfnamefont {R.~A.}\ \bibnamefont
  {{M}arcus}},\ }\href@noop {} {\bibfield  {journal} {\bibinfo  {journal} {The
  Journal of Physical Chemistry}\ }\textbf {\bibinfo {volume} {72}},\ \bibinfo
  {pages} {891} (\bibinfo {year} {1968})}\BibitemShut {NoStop}%
\bibitem [{\citenamefont {{F}rank {J}ensen}(2007)}]{Jensen2007}%
  \BibitemOpen
  \bibfield  {author} {\bibinfo {author} {\bibnamefont {{F}rank {J}ensen}},\
  }\href@noop {} {\emph {\bibinfo {title} {{I}ntroduction to {C}omputational
  {C}hemistry}}}\ (\bibinfo  {publisher} {John Wiley \& Sons},\ \bibinfo
  {address} {New York},\ \bibinfo {year} {2007})\BibitemShut {NoStop}%
\bibitem [{\citenamefont {{R}oy}, \citenamefont {{G}oedecker},\ and\
  \citenamefont {{H}ellmann}(2008)}]{Roy2008}%
  \BibitemOpen
  \bibfield  {author} {\bibinfo {author} {\bibfnamefont {S.}~\bibnamefont
  {{R}oy}}, \bibinfo {author} {\bibfnamefont {S.}~\bibnamefont {{G}oedecker}},
  \ and\ \bibinfo {author} {\bibfnamefont {V.}~\bibnamefont {{H}ellmann}},\
  }\href@noop {} {\bibfield  {journal} {\bibinfo  {journal} {Physical Review
  E}\ }\textbf {\bibinfo {volume} {77}},\ \bibinfo {pages} {056707} (\bibinfo
  {year} {2008})}\BibitemShut {NoStop}%
\bibitem [{\citenamefont {{S}icher}, \citenamefont {{M}ohr},\ and\
  \citenamefont {{G}oedecker}(2011)}]{Sicher2011}%
  \BibitemOpen
  \bibfield  {author} {\bibinfo {author} {\bibfnamefont {M.}~\bibnamefont
  {{S}icher}}, \bibinfo {author} {\bibfnamefont {S.}~\bibnamefont {{M}ohr}}, \
  and\ \bibinfo {author} {\bibfnamefont {S.}~\bibnamefont {{G}oedecker}},\
  }\href@noop {} {\bibfield  {journal} {\bibinfo  {journal} {The Journal of
  Chemical Physics}\ }\textbf {\bibinfo {volume} {134}},\ \bibinfo {pages}
  {044106} (\bibinfo {year} {2011})}\BibitemShut {NoStop}%
\bibitem [{\citenamefont {{B}orn}\ and\ \citenamefont
  {{M}ayer}(1932)}]{Born1932}%
  \BibitemOpen
  \bibfield  {author} {\bibinfo {author} {\bibfnamefont {M.}~\bibnamefont
  {{B}orn}}\ and\ \bibinfo {author} {\bibfnamefont {J.}~\bibnamefont
  {{M}ayer}},\ }\href@noop {} {\bibfield  {journal} {\bibinfo  {journal}
  {Zeitschrift f\"{u}r Physik}\ }\textbf {\bibinfo {volume} {75}},\ \bibinfo
  {pages} {1} (\bibinfo {year} {1932})}\BibitemShut {NoStop}%
\bibitem [{\citenamefont {{M}ayer}(1933)}]{Mayer1933}%
  \BibitemOpen
  \bibfield  {author} {\bibinfo {author} {\bibfnamefont {J.~E.}\ \bibnamefont
  {{M}ayer}},\ }\href@noop {} {\bibfield  {journal} {\bibinfo  {journal} {The
  Journal of Chemical Physics}\ }\textbf {\bibinfo {volume} {1}},\ \bibinfo
  {pages} {270} (\bibinfo {year} {1933})}\BibitemShut {NoStop}%
\bibitem [{\citenamefont {{H}uggins}\ and\ \citenamefont
  {{M}ayer}(1933)}]{Huggins1933}%
  \BibitemOpen
  \bibfield  {author} {\bibinfo {author} {\bibfnamefont {M.~L.}\ \bibnamefont
  {{H}uggins}}\ and\ \bibinfo {author} {\bibfnamefont {J.~E.}\ \bibnamefont
  {{M}ayer}},\ }\href@noop {} {\bibfield  {journal} {\bibinfo  {journal} {The
  Journal of Chemical Physics}\ }\textbf {\bibinfo {volume} {1}},\ \bibinfo
  {pages} {643} (\bibinfo {year} {1933})}\BibitemShut {NoStop}%
\bibitem [{\citenamefont {{F}umi}\ and\ \citenamefont
  {{T}osi}(1964)}]{Fumi1964}%
  \BibitemOpen
  \bibfield  {author} {\bibinfo {author} {\bibfnamefont {F.}~\bibnamefont
  {{F}umi}}\ and\ \bibinfo {author} {\bibfnamefont {M.}~\bibnamefont
  {{T}osi}},\ }\href@noop {} {\bibfield  {journal} {\bibinfo  {journal}
  {Journal of Physics and Chemistry of Solids}\ }\textbf {\bibinfo {volume}
  {25}},\ \bibinfo {pages} {31 } (\bibinfo {year} {1964})}\BibitemShut
  {NoStop}%
\bibitem [{\citenamefont {{T}osi}\ and\ \citenamefont
  {{F}umi}(1964)}]{Tosi1964}%
  \BibitemOpen
  \bibfield  {author} {\bibinfo {author} {\bibfnamefont {M.}~\bibnamefont
  {{T}osi}}\ and\ \bibinfo {author} {\bibfnamefont {F.}~\bibnamefont
  {{F}umi}},\ }\href@noop {} {\bibfield  {journal} {\bibinfo  {journal}
  {Journal of Physics and Chemistry of Solids}\ }\textbf {\bibinfo {volume}
  {25}},\ \bibinfo {pages} {45 } (\bibinfo {year} {1964})}\BibitemShut
  {NoStop}%
\bibitem [{\citenamefont {{P}erdew}, \citenamefont {{B}urke},\ and\
  \citenamefont {{E}rnzerhof}(1996)}]{Perdew1996}%
  \BibitemOpen
  \bibfield  {author} {\bibinfo {author} {\bibfnamefont {J.~P.}\ \bibnamefont
  {{P}erdew}}, \bibinfo {author} {\bibfnamefont {K.}~\bibnamefont {{B}urke}}, \
  and\ \bibinfo {author} {\bibfnamefont {M.}~\bibnamefont {{E}rnzerhof}},\
  }\href {http://link.aps.org/doi/10.1103/PhysRevLett.77.3865} {\bibfield
  {journal} {\bibinfo  {journal} {Physical Review Letters}\ }\textbf {\bibinfo
  {volume} {77}},\ \bibinfo {pages} {3865} (\bibinfo {year}
  {1996})}\BibitemShut {NoStop}%
\bibitem [{\citenamefont {{K}ohn}\ and\ \citenamefont
  {{S}ham}(1965)}]{Kohn1965}%
  \BibitemOpen
  \bibfield  {author} {\bibinfo {author} {\bibfnamefont {W.}~\bibnamefont
  {{K}ohn}}\ and\ \bibinfo {author} {\bibfnamefont {L.~J.}\ \bibnamefont
  {{S}ham}},\ }\href@noop {} {\bibfield  {journal} {\bibinfo  {journal}
  {Physical Review}\ }\textbf {\bibinfo {volume} {140}},\ \bibinfo {pages}
  {A1133} (\bibinfo {year} {1965})}\BibitemShut {NoStop}%
\bibitem [{\citenamefont {{W}eitao}\ and\ \citenamefont
  {{Y}ang}(1994)}]{ParrYang94}%
  \BibitemOpen
  \bibfield  {author} {\bibinfo {author} {\bibfnamefont {R.~G.~P.}\
  \bibnamefont {{W}eitao}}\ and\ \bibinfo {author} {\bibnamefont {{Y}ang}},\
  }\href@noop {} {\emph {\bibinfo {title} {{D}ensity-{F}unctional {T}heory of
  {A}toms and {M}olecules}}}\ (\bibinfo  {publisher} {Oxford University
  Press},\ \bibinfo {address} {Oxford},\ \bibinfo {year} {1994})\BibitemShut
  {NoStop}%
\bibitem [{\citenamefont {{A}dams}\ and\ \citenamefont
  {{M}c{D}onald}(1975)}]{Adams1975}%
  \BibitemOpen
  \bibfield  {author} {\bibinfo {author} {\bibfnamefont {D.~J.}\ \bibnamefont
  {{A}dams}}\ and\ \bibinfo {author} {\bibfnamefont {I.~R.}\ \bibnamefont
  {{M}c{D}onald}},\ }\href@noop {} {\bibfield  {journal} {\bibinfo  {journal}
  {Journal of Physics C: Solid State Physics}\ }\textbf {\bibinfo {volume}
  {8}},\ \bibinfo {pages} {2198} (\bibinfo {year} {1975})}\BibitemShut
  {NoStop}%
\bibitem [{\citenamefont {{J}ones}(1924)}]{Lennard-Jones1924}%
  \BibitemOpen
  \bibfield  {author} {\bibinfo {author} {\bibfnamefont {J.~E.}\ \bibnamefont
  {{J}ones}},\ }\href@noop {} {\bibfield  {journal} {\bibinfo  {journal}
  {Proceedings of the Royal Society of London A}\ }\textbf {\bibinfo {volume}
  {106}},\ \bibinfo {pages} {463} (\bibinfo {year} {1924})}\BibitemShut
  {NoStop}%
\bibitem [{\citenamefont {{J}ones}\ and\ \citenamefont
  {{I}ngham}(1925)}]{Lennard-Jones1925}%
  \BibitemOpen
  \bibfield  {author} {\bibinfo {author} {\bibfnamefont {J.~E.}\ \bibnamefont
  {{J}ones}}\ and\ \bibinfo {author} {\bibfnamefont {A.~E.}\ \bibnamefont
  {{I}ngham}},\ }\href@noop {} {\bibfield  {journal} {\bibinfo  {journal}
  {Proceedings of the Royal Society of London A}\ }\textbf {\bibinfo {volume}
  {107}},\ \bibinfo {pages} {636} (\bibinfo {year} {1925})}\BibitemShut
  {NoStop}%
\bibitem [{\citenamefont {{S}chaefer}\ \emph
  {et~al.}(2014{\natexlab{b}})\citenamefont {{S}chaefer}, \citenamefont
  {{P}al}, \citenamefont {{K}hetrapal}, \citenamefont {{A}msler}, \citenamefont
  {{S}adeghi}, \citenamefont {{B}lum}, \citenamefont {{Z}eng}, \citenamefont
  {{G}oedecker},\ and\ \citenamefont {{W}ang}}]{Schaefer2014a}%
  \BibitemOpen
  \bibfield  {author} {\bibinfo {author} {\bibfnamefont {B.}~\bibnamefont
  {{S}chaefer}}, \bibinfo {author} {\bibfnamefont {R.}~\bibnamefont {{P}al}},
  \bibinfo {author} {\bibfnamefont {N.~S.}\ \bibnamefont {{K}hetrapal}},
  \bibinfo {author} {\bibfnamefont {M.}~\bibnamefont {{A}msler}}, \bibinfo
  {author} {\bibfnamefont {A.}~\bibnamefont {{S}adeghi}}, \bibinfo {author}
  {\bibfnamefont {V.}~\bibnamefont {{B}lum}}, \bibinfo {author} {\bibfnamefont
  {X.~C.}\ \bibnamefont {{Z}eng}}, \bibinfo {author} {\bibfnamefont
  {S.}~\bibnamefont {{G}oedecker}}, \ and\ \bibinfo {author} {\bibfnamefont
  {L.-S.}\ \bibnamefont {{W}ang}},\ }\href@noop {} {\bibfield  {journal}
  {\bibinfo  {journal} {ACS Nano}\ }\textbf {\bibinfo {volume} {8}},\ \bibinfo
  {pages} {7413} (\bibinfo {year} {2014}{\natexlab{b}})}\BibitemShut {NoStop}%
\bibitem [{\citenamefont {{L}evenberg}(1944)}]{Levenberg1944}%
  \BibitemOpen
  \bibfield  {author} {\bibinfo {author} {\bibfnamefont {K.}~\bibnamefont
  {{L}evenberg}},\ }\href@noop {} {\bibfield  {journal} {\bibinfo  {journal}
  {Quarterly of Applied Mathematics}\ }\textbf {\bibinfo {volume} {II}},\
  \bibinfo {pages} {164} (\bibinfo {year} {1944})}\BibitemShut {NoStop}%
\bibitem [{\citenamefont {{D}onald {W}.~{M}arquardt}(1963)}]{Marquardt1963}%
  \BibitemOpen
  \bibfield  {author} {\bibinfo {author} {\bibnamefont {{D}onald
  {W}.~{M}arquardt}},\ }\href@noop {} {\bibfield  {journal} {\bibinfo
  {journal} {Journal of the Society for Industrial and Applied Mathematics}\
  }\textbf {\bibinfo {volume} {11}},\ \bibinfo {pages} {431} (\bibinfo {year}
  {1963})}\BibitemShut {NoStop}%
\bibitem [{\citenamefont {{T}homas {W}illiams}, \citenamefont {{C}olin
  {K}elley},\ and\ \citenamefont {{many others}}(2014)}]{gnuplot}%
  \BibitemOpen
  \bibfield  {author} {\bibinfo {author} {\bibnamefont {{T}homas {W}illiams}},
  \bibinfo {author} {\bibnamefont {{C}olin {K}elley}}, \ and\ \bibinfo {author}
  {\bibnamefont {{many others}}},\ }\href@noop {} {\enquote {\bibinfo {title}
  {{G}nuplot 4.6: {A}n {I}nteractive {P}lotting {P}rogram},}\ }\bibinfo
  {howpublished} {\url{http://gnuplot.sourceforge.net}} (\bibinfo {year}
  {2014})\BibitemShut {NoStop}%
\bibitem [{\citenamefont {{D}ellago}(2007)}]{Dellago2007}%
  \BibitemOpen
  \bibfield  {author} {\bibinfo {author} {\bibfnamefont {C.}~\bibnamefont
  {{D}ellago}},\ }in\ \href@noop {} {\emph {\bibinfo {booktitle} {Free Energy
  Calculations}}},\ \bibinfo {series} {Springer Series in Chemical Physics},
  Vol.~\bibinfo {volume} {86},\ \bibinfo {editor} {edited by\ \bibinfo {editor}
  {\bibfnamefont {C.}~\bibnamefont {Chipot}}\ and\ \bibinfo {editor}
  {\bibfnamefont {A.}~\bibnamefont {Pohorille}}}\ (\bibinfo  {publisher}
  {Springer Berlin Heidelberg},\ \bibinfo {year} {2007})\ pp.\ \bibinfo {pages}
  {249--276}\BibitemShut {NoStop}%
\bibitem [{\citenamefont {{D}oye}, \citenamefont {{M}iller},\ and\
  \citenamefont {{W}ales}(1999{\natexlab{a}})}]{Doye1999}%
  \BibitemOpen
  \bibfield  {author} {\bibinfo {author} {\bibfnamefont {J.~P.~K.}\
  \bibnamefont {{D}oye}}, \bibinfo {author} {\bibfnamefont {M.~A.}\
  \bibnamefont {{M}iller}}, \ and\ \bibinfo {author} {\bibfnamefont {D.~J.}\
  \bibnamefont {{W}ales}},\ }\href@noop {} {\bibfield  {journal} {\bibinfo
  {journal} {The Journal of Chemical Physics}\ }\textbf {\bibinfo {volume}
  {110}},\ \bibinfo {pages} {6896} (\bibinfo {year}
  {1999}{\natexlab{a}})}\BibitemShut {NoStop}%
\bibitem [{\citenamefont {{D}oye}, \citenamefont {{M}iller},\ and\
  \citenamefont {{W}ales}(1999{\natexlab{b}})}]{Doye1999-2}%
  \BibitemOpen
  \bibfield  {author} {\bibinfo {author} {\bibfnamefont {J.~P.~K.}\
  \bibnamefont {{D}oye}}, \bibinfo {author} {\bibfnamefont {M.~A.}\
  \bibnamefont {{M}iller}}, \ and\ \bibinfo {author} {\bibfnamefont {D.~J.}\
  \bibnamefont {{W}ales}},\ }\href@noop {} {\bibfield  {journal} {\bibinfo
  {journal} {The Journal of Chemical Physics}\ }\textbf {\bibinfo {volume}
  {111}},\ \bibinfo {pages} {8417} (\bibinfo {year}
  {1999}{\natexlab{b}})}\BibitemShut {NoStop}%
\bibitem [{\citenamefont {{R}ahman}(1964)}]{Rahman1964}%
  \BibitemOpen
  \bibfield  {author} {\bibinfo {author} {\bibfnamefont {A.}~\bibnamefont
  {{R}ahman}},\ }\href {http://link.aps.org/doi/10.1103/PhysRev.136.A405}
  {\bibfield  {journal} {\bibinfo  {journal} {Physical Review}\ }\textbf
  {\bibinfo {volume} {136}},\ \bibinfo {pages} {A405} (\bibinfo {year}
  {1964})}\BibitemShut {NoStop}%
\bibitem [{\citenamefont {{R}owley}, \citenamefont {{N}icholson},\ and\
  \citenamefont {{P}arsonage}(1975)}]{Rowley1975}%
  \BibitemOpen
  \bibfield  {author} {\bibinfo {author} {\bibfnamefont {L.}~\bibnamefont
  {{R}owley}}, \bibinfo {author} {\bibfnamefont {D.}~\bibnamefont
  {{N}icholson}}, \ and\ \bibinfo {author} {\bibfnamefont {N.}~\bibnamefont
  {{P}arsonage}},\ }\href@noop {} {\bibfield  {journal} {\bibinfo  {journal}
  {Journal of Computational Physics}\ }\textbf {\bibinfo {volume} {17}},\
  \bibinfo {pages} {401} (\bibinfo {year} {1975})}\BibitemShut {NoStop}%
\bibitem [{\citenamefont {{W}hite}(1999)}]{White1999}%
  \BibitemOpen
  \bibfield  {author} {\bibinfo {author} {\bibfnamefont {J.~A.}\ \bibnamefont
  {{W}hite}},\ }\href@noop {} {\bibfield  {journal} {\bibinfo  {journal} {The
  Journal of Chemical Physics}\ }\textbf {\bibinfo {volume} {111}},\ \bibinfo
  {pages} {9352} (\bibinfo {year} {1999})}\BibitemShut {NoStop}%
\end{thebibliography}%

\end{document}